\documentclass[12pt, nofootinbib]{revtex4}
\usepackage{graphicx}
\usepackage{amssymb}
\usepackage{axodraw}
\usepackage{slashed}
\usepackage{subfigure}

\newcommand{\be}{\begin{equation}}
\newcommand{\ee}{\end{equation}}
\newcommand{\nn}{\nonumber}
\newcommand{\bs}{\bigskip}

\begin{document}

\title{About Fermion Hierarchies from Doubly Warped Extra Dimensions.}
\author{R. Lawrance}
\affiliation{Department of Physics, College of Science, Swansea University, Singleton Park\\ Swansea, UK\\
605879@swansea.ac.uk}

\begin{abstract}

We consider fermions propagating in the bulk of the geometry found by deforming $AdS_5$ via the back reaction of a scalar field upon the metric. This space is AdS for $r$ asymptotically large (in the UV) but goes through a transition at a point $r=r_*$, into another AdS space with different curvature in the IR. Masses are generated for these fermions via electroweak symmetry breaking, by coupling them to a VEV on the IR boundary. We calculate the mass spectrum in four dimensions, comparing approximate results and results found by solving the full system of bulk equations and boundary conditions. We consider the effect on the mass of the light modes of various parameters, including the curvature of the space in the region $r<r_*$. This information is then used to reproduce the mass hierarchy between the top and bottom. By assuming universality of the gauge coupling, we find bounds on the allowed bulk masses of the right--handed fermion fields. We look for solutions that satisfy these bounds in a number of different scenarios and find that, for given choices of the other parameters in this model, the IR curvature has a significant influence on whether these bounds can be satisfied or not.
\end{abstract}

\maketitle

\tableofcontents

\section{Introduction}

One of the mysteries in our understanding of the standard model is the nature of the mechanism responsible for generating the large mass hierarchies between the standard--model fermions. One approach to understanding this problem, in the context of strongly coupled models of electroweak symmetry breaking, is sketched out by extended technicolor \cite{ETC, ETC2}, which relies on various symmetries to suppress the masses of some (or all) of the standard--model fermions. This is not the only possible approach to explaining the fermion mass hierarchy. It has also been suggested that, by allowing fermions to propagate in an extra dimension, suppression of their mass is possible without introducing any extra symmetries.

\bs

Using extra dimensions to explain the mass hierarchy of the standard--model fermions was first proposed by Arkani-Hamed and Schmaltz \cite{arkani-hamed}. The basic idea was to localize the fermions at different points in a flat extra dimension by coupling them to scalar fields which posses a kink--shaped bulk profile. The mass of a fermion would then be proportional to the overlap of the wavefunctions of the left-- and right--handed fields, and suppression of the mass was possible due to the fact that the two fields were localized differently.

\bs

Theories of warped extra dimensions, particularly those where the extra dimension is AdS or asymptotically AdS (such as the simple model of Randall and Sundrum \cite{RS1}), have also attracted a lot of attention due to the important role they play in the AdS/CFT correspondence \cite{AdS/CFT, AdS/CFT2, AdS/CFT3, AdS/CFT4} and the more general notion of a gauge--gravity duality where the bulk theory, which is weakly coupled and includes gravity, is related to a strongly coupled theory living on the boundary. Much work has been done in developing the formalism for treating bulk fermions in AdS \cite{gherg, ConPom, csaki et al, GIM, dim} and computing the mass spectrum in various models (e.g. supersymmetry in AdS) \cite{models, models2, models3, nomburd, leptons, BDR, car}. Other aspects of the physics of fermions in warped extra dimensions has also been considered in the literature, including calculation of electroweak precision parameters \cite{S1, S2} flavour physics \cite{o3, o4, o5, fl1, fl2}, baryon physics \cite{o6, o7}, the anomalous magnetic moment of the muon \cite{o8} and neutrino mixing \cite{o2}.

\bs

To compute the mass spectrum, first one wants to construct a model in which chirality is recovered in the four dimensional theory. In this case the fermions are massless but a mass can be generated for them via electroweak symmetry breaking, by coupling the fermions to a VEV placed on one of the boundaries of the space (either in the UV or the IR)\footnote{Note that the approach taken in \cite{BDR, car} is somewhat different. In these models a fourth generation of bulk fermions is introduced and assumed to condense. This dynamically generates the VEV responsible for fermion masses.}. The mass of the fermion is then determined by the bulk dynamics i.e. how the fermion is localized in the extra dimension. One of the important parameters in determining the bulk dynamics of the fermions propagating in warped extra dimensions is the curvature of the space. In AdS (where the curvature is constant) its role is somewhat trivial. However this need not be the case for asymptotically AdS geometries such as the case where AdS is deformed by the backreaction of a scalar field upon the metric. 

\bs

A toy model of this class is developed in \cite{DE and MP, RL and MP} in the context of a holographic model of Technicolor \cite{TC, TC2, TC3}. In this model the scalar field has a kink--shaped bulk profile and leads to a space that is AdS in the UV but, moving into the IR, undergoes a transition to an AdS space of a different curvature, with this transition taking place around the position of the centre of the kink in the scalar profile. We will consider the dynamics of bulk fermions propagating in AdS and the space described here, which we refer to as the deformed background. Our aim is to explore the dependence of the four dimensional mass of the fermions on the various parameters in the model, paying particular attention to the role of the curvature in this deformed background. We will then use this to build a simple model which reproduces the mass hierarchy of the standard--model fermions, in particular the top and bottom quarks.

\bs

The paper is organised as follows: in section 2 we review elements from \cite{RL and MP}, explaining the generation and nature of the deformed background. Section 3 will consider approximate and exact calculations of the fermion mass spectrum, both in AdS and the deformed background. We will also comment on the dependence of the solutions on the various parameters and their importance. In section 4 we consider the gauge coupling of the fermions to the $Z$ boson and use this to place indicative phenomenological bounds on our model, while in section 5 we consider how the $\hat{S}$ parameter is effected by such considerations. Section 6 consists of a discussion of the masses of the top and bottom quarks, while section 7 contains our conclusions.

\section{Setup}

\subsection{Geometry}

Consider the five dimensional space-time, defined by the metric
\be
ds^2=g_{\bar{M}\bar{N}}dx^{\bar{M}}dx^{\bar{N}}=e^{2A(r)}\eta_{\mu\nu}dx^\mu dx^\nu +dr^2,
\ee
where a warp factor of the form $A(r)=\kappa r$ describes an AdS space of curvature $\kappa$. We use a metric with signature $(-,+,+,+,+)$ and use lower--case greek indices to label curved 4D coordinates and barred capital latin indices to label curved 5D coordinates. For any point on our curved manifold, a flat tangent space can be found
\be\label{curvedmetric}
g_{\bar{M}\bar{N}}=e_{\bar{M}}^{\,\,\,\,\,M} e_{\bar{N}}^{\,\,\,\,\,N}\eta_{MN}
=\left(\begin{array}{cc}
e_\mu^{\,\,\,m} e_\nu^{\,\,\,n} \eta_{mn}&\\
&e_r^{\,\,\,5} e_r^{\,\,\,5} \eta_{55}
\end{array}\right)\,,
\ee
where we use capital latin indices (no bar) to label 5D flat coordinates and lower case latin indices to label 4D flat coordinates (we also use a lower case $r$ to label the fifth coordinate in curved space and a $5$ to label the fifth coordinate in the tangent space). Eq.~(\ref{curvedmetric}) then defines the vielbein $e_{\bar{M}}^{\,\,\,\,\,M}$, which describes the relationship between the two spaces. In the basis in which $g_{\bar{M}\bar{N}}$ is diagonal, we write
\be\label{vielbein}
e_{\bar{M}}^{\,\,\,\,\,M}=\left(\begin{array}{cc}
e^{A(r)}\delta_\mu^{\,\,\,m}&\\
&1
\end{array}\right)\,.
\ee
To this space we add an IR boundary at $r=r_1$, to act as an IR cut off and a UV boundary at $r=r_2$. We set $r_1=0$ in all subsequent calculations.

\subsection{Scalar Background}

Given the geometry described in the previous subsection, we couple to gravity a $\sigma$--model consisting of a set of scalar fields $\Phi^a$ with internal $\sigma$--model metric $G_{ab}=\delta_{ab}$, such that the $\sigma$--model connection $\mathcal{G}^c_{ab}=0$. The action is
\begin{eqnarray}\label{scalaraction}
\mathcal{S}&=&\int d^4x dr \sqrt{-g}\Theta\left(\frac{R}{4}+\mathcal{L}_5\right)+\sqrt{-\tilde{g}}\delta(r-r_1)\left(\frac{K}{2}+\mathcal{L}_1\right)\nonumber\\&&-\sqrt{-\tilde{g}}\delta(r-r_2)\left(\frac{K}{2}+\mathcal{L}_2\right)\,,
\end{eqnarray}
where $\tilde{g}_{\mu\nu}$ is the induced boundary metric, $R$ is the Ricci scalar and $K$ is the extrinsic curvature of the boundary hyper-surface, defined by
\begin{eqnarray}
K_{\mu\nu}&=&\nabla_{\mu} N_{\nu}\,,\,\,\,\,
K\,=\,\tilde{g}^{\mu\nu}K_{\mu\nu}.
\end{eqnarray}
$N_{\nu}$ is an orthonormal vector to the surface, and
\be
\mathcal{L}_5=-\frac{1}{2}g^{\bar{M}\bar{N}}\partial_{\bar{M}}\Phi^a\partial_{\bar{N}}\Phi_a-V(\Phi^a)\,,\,
\mathcal{L}_1\,=\,-\lambda_1(\Phi^a)\,,\,
\mathcal{L}_2\,=\,-\lambda_2(\Phi^a)\,,
\ee
where $V(\Phi^a)$ is a bulk potential and the $\lambda_i(\Phi^a)$ are localized potentials on the 4D boundaries. Varying Eq.~(\ref{scalaraction}) with respect to the metric yields the Einstein equations
\begin{eqnarray}\label{EE1}
6\left(A^\prime\right)^2+3A^{\prime\prime}+\Phi^{\prime\,a}\Phi^\prime_a+2V&=&0\,,\\
\label{EE2}
6\left(A^\prime\right)^2-\Phi^{\prime\,a}\Phi^\prime_a+2V&=&0\,,
\end{eqnarray}
while varying with respect to the scalar fields gives the equations of motion and boundary conditions for the scalars. Imposing 4D Poincar\'e invariance on the scalars, we find
\be
\bar{\Phi}^{\prime\prime\,a}+4A^\prime\bar{\Phi}^{\prime\,a}-\partial_{\Phi_a} V=0.
\ee
We also have the boundary conditions
\begin{eqnarray}
\bar{\Phi}^{\prime\,a}|_{r_i}&=&\partial_{\Phi_a}\lambda_i|_{r_i}\,,\\
A^\prime|_{r_i}&=&-\frac{2}{3}\lambda_i|_{r_i}\,,
\end{eqnarray}
where $\bar{\Phi}^a$ is the classical solution. These boundary conditions constrain the form of the $\lambda_i$
\be
\lambda_i=-\frac{3}{2}A^{\prime}|_{r_i}+\frac{}{}\bar{\Phi}^{a\,\prime}|_{r_i}\left(\Phi_a-\Phi_a(r_i)\right)+\cdots\,.
\ee
If the potential $V(\Phi^a)$ can be written in terms of a superpotential
\be
V=\frac{1}{2}(\partial_{\Phi_a} W)^2-\frac{4}{3}W^2
\ee
then it is possible to expand the $\lambda_i$ in terms of the superpotential
\be
\lambda_i=W(\Phi(r_i))+\partial_{\Phi^a}W(\Phi(r_i))(\Phi^a-\Phi^a(r_i))+\cdots
\ee
Therefore, at leading order we have
\begin{equation}\label{eqn:Aprime}
A^\prime=-\frac{2}{3}W,
\end{equation}
and
\begin{equation}\label{Phiprime}
\bar{\Phi}^{\prime\,a}=\partial_{\Phi_a} W,
\end{equation}
and it follows that solutions to Eq.~(\ref{eqn:Aprime}) and Eq.~(\ref{Phiprime}) are also solutions to the equations of motion and Einstein equations. As such, one only needs to solve Eq.~(\ref{eqn:Aprime}) and Eq.~(\ref{Phiprime}) to yield the background.

\bs

We are ultimately interested in the dynamics of fermions probing a deformed background. To this end we focus on the model introduced in \cite{DE and MP} and developed further in \cite{RL and MP} where the background is generated by a single scalar with a superpotential of the form
\be
W=-\frac{3}{2}-\frac{\Delta}{2}\Phi^2+\frac{\Delta}{3\Phi_I}\Phi^3\,,
\ee
where $\Delta$ and $\Phi_I$ are free parameters. Solving Eq.~(\ref{Phiprime}) for this choice of superpotential gives the classical solution
\begin{equation}\label{phicl}
\bar{\Phi}=\frac{\Phi_I}{1+e^{\Delta(r-r_*)}}\,,
\end{equation}
while solving Eq.~(\ref{eqn:Aprime}) gives the warp factor
\begin{eqnarray}
A(r)&=&r+\frac{\Phi ^{2}_I}{9} \left(\frac{}{}\Delta r-\ln\left(e^{\Delta 
\left(r-r_*\right)}+1\right)\right.\nn\\&&\left.+\frac{1}{e^{\Delta 
\left(r-r_*\right)}+1}-\frac{1}{\left(e^{\Delta 
\left(r-r_*\right)}+1\right){}^2}\right)\,,
\end{eqnarray}
where $r_*$ is an integration constant, which arises when solving Eq.~(\ref{Phiprime}). Note that another integration constant is found when solving Eq.~(\ref{eqn:Aprime}), which is determined by setting $A(0)=0$. The classical solutions for $\bar{\Phi}(r)$ and $A(r)$ are shown in Figure 1 for $\Phi_I=1$, $\Delta=5$ and $r_*=5$. Note that the warp factor describes a bulk geometry which is approximately AdS in the regions $r<r_*$ and $r>r_*$ but in which the curvature changes smoothly around this point. This means that the complicated expression above can be well approximated by
\be\label{Aapprox}
A(r)\simeq\left\{\begin{array}{l} \kappa_0 r\,,\,\,r<r_*\nn\\
\kappa_1 r+(\kappa_0-\kappa_1)r_*\,,\,\,r>r_*\nn\\
\end{array}\right.\,,
\ee
where $\kappa_i$ are the curvatures of each region, and are given by
\begin{eqnarray}
\kappa_0&=&1+\frac{\Delta\Phi_I^2e^{2\Delta r_*}\left(e^{\Delta r_*}+3\right)}{9\left(1+e^{\Delta r_*}\right)^3}\,,\nn\\
\kappa_1&=&1\,.
\end{eqnarray}
The validity of this approximation is dependent on the sharpness of the kink in the scalar profile. This is controlled by the parameter $\Delta$, therefore $\Delta$ should be taken sufficiently large. What is sufficiently large is ultimately determined by the sensitivity of subsequent calculations to this approximation. In the context of this paper, where we are interested in the calculation of fermion spectra, $\Delta\geq 1$ is sufficient (actually, $\Delta$ less than, but very close to, one may also be sufficient). Also note that, for $\Delta r_*>>1$, $\kappa_0$ can be approximated by $\kappa_0=\kappa_1+\delta\kappa$ where
\be\label{dk}
\delta\kappa\rightarrow\frac{\Delta\Phi_I^2}{9}\,.
\ee

\begin{figure}[t]
\begin{center}
\subfigure{\includegraphics[scale=0.7]{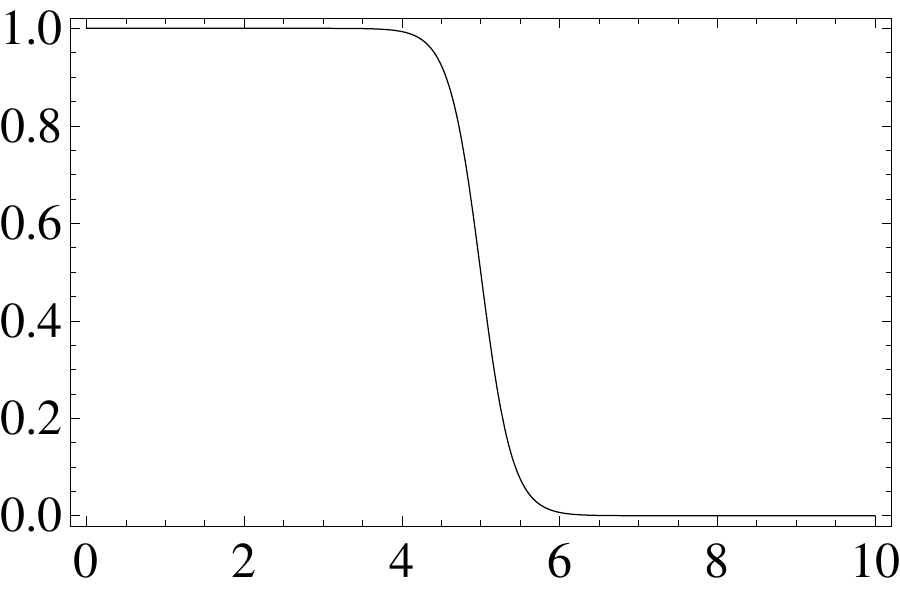}
\Text(-190,110)[]{\small{$\bar{\Phi}$}}
\Text(-20,0)[]{\small{$r$}}}
\subfigure{\includegraphics[scale=0.7]{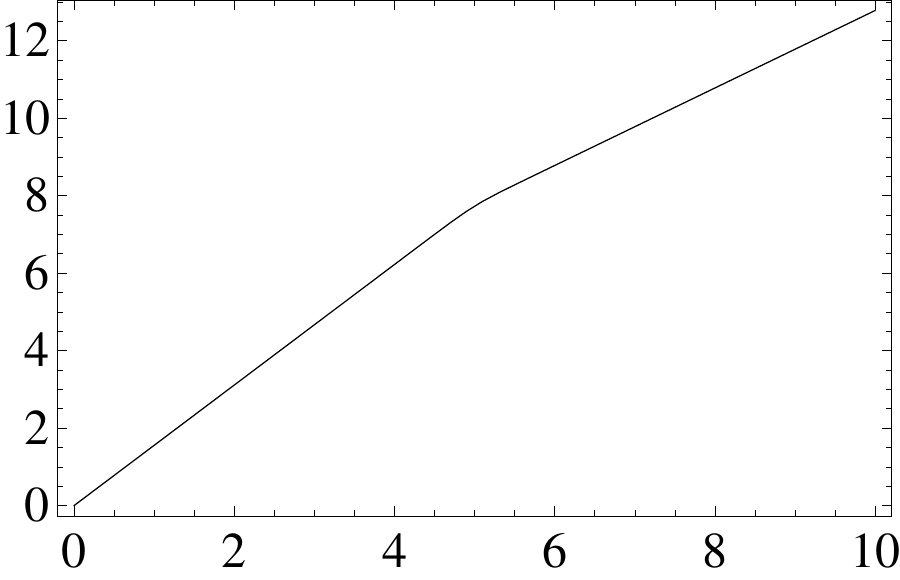}
\Text(10,110)[]{\small{$A$}}
\Text(-20,0)[]{\small{$r$}}}
\end{center}\caption{Left panel: plot of $\bar{\Phi}$ against $r$ for $\Phi_I=1$, $\Delta=5$ and $r_*=5$. Note that $\bar{\Phi}$ is approximately constant in the two regions $r<r_*$ and $r>r_*$. Right panel: plot of $A(r)$ against $r$ for $\Phi_I=1$, $\Delta=5$ and $r_*=5$. Note that $A(r)$ approximately linear in the two regions $r<r_*$ and $r>r_*$.}
\label{phibarAplot}
\end{figure}

\section{General Results}

\subsection{Chiral Fermions}

We consider fermions allowed to propagate in the bulk, in probe approximation, using the formalism developed in \cite{ConPom}. These fermions are introduced via the 5D action
\be\label{fermionaction}
\mathcal{S}=\int d^4x\int_{r_1}^{r_2} dr \sqrt{-g}\left(i\bar{\Psi}^ie^{\bar{M}}_{\,\,\,\,A}\Gamma^A D_{\bar{M}}\Psi^i-M_i\bar{\Psi}^i\Psi^i\right)\,,
\ee
where $M$ is a bulk mass. We indicate with $\Gamma^A$ the $\gamma$ matrices in five dimensions, with $D_{\bar{M}}$ the covariant derivative
\be
D_{\bar{M}}=\partial_{\bar{M}}+\frac{1}{8}\omega_{\bar{M}BC}\left[\Gamma^B,\Gamma^C\right]\,,
\ee
and $\omega_{\bar{M}BC}$ is the spin-connection, which can be expressed in terms of the torsion $T^A_{\,\,\,\,BC}$ as
\begin{eqnarray}\label{sctorsion}
\omega_{\bar{M}}^{\,\,\,\,MN}&=&\frac{1}{2}e^{\,\,\,\,\,Q}_{\bar{M}}\left(\eta_{QA}\eta^{MB}\eta^{NC}-\delta^{\,\,\,\,M}_A\eta^{NB}\delta_Q^{\,\,\,\,C}-\delta_A^{\,\,\,\,N}\delta_Q^{\,\,\,\,B}\eta^{MC}\right)T^A_{\,\,\,\,BC}\,,\\
T^A_{\,\,\,\,BC}&=&\left(e_{\,\,\,\,B}^{\bar{M}}e_{\,\,\,\,C}^{\bar{P}}-e_{\,\,\,\,C}^{\bar{M}}e_{\,\,\,\,B}^{\bar{P}}\right)\partial_{\bar{P}}e_{\,\,\,\,\bar{M}}^A\,.
\end{eqnarray}
Solving Eq.~(\ref{sctorsion}), the only non-zero components of the (antisymmetric) spin-connection are
\be
\omega_\mu^{\,\,\,m5}=A^\prime(r) e^A(r) \delta_\mu^{\,\,\,m}\,.
\ee
Note that, since the matrices $\Gamma^A$ carry flat space indices, these reduce to the 4D Dirac gamma matrices $\gamma^\mu$, plus $\Gamma^5=-i\gamma^5$.

\bs

Now we decompose the fermion $\Psi=\psi_L+\psi_R$ into left-- and right--handed components (we drop the field index $i$, this will be reintroduced later if necessary), where $\psi_{L,R}=\frac{1}{2}(I_4\mp\gamma^5)\Psi$. Performing a Fourier transformation on the 4D coordinates and  applying the variational principle yields the bulk equations
\begin{eqnarray}\label{EoM1}
-e^{-A(r)}\slashed{p}\,\psi_{R}+\partial_r\psi_{L}+2A^\prime(r)\psi_{L} + M\psi_{L}&=&0\,,\\
e^{-A(r)}\slashed{p}\,\psi_{L}+\partial_r\psi_{R}+2A^\prime(r)\psi_{R} - M\psi_{R}&=&0\,.
\end{eqnarray}
Decomposing the fermions as
\be\label{dec}
\psi_{L,R}(p,r)=\frac{f_{L,R}(p,r)}{f_{L,R}(p,r_2)}\psi^0_{L,R}(p)\,,
\ee
it is possible to show that the functions $f_{L,R}(p,r)$ satisfy the first--order coupled differential equations
\begin{eqnarray}\label{eqnf}
 p\,e^{-A(r)}f_{R}(p,r)&=&\partial_r f_{L}(p,r)+2A^\prime(r)f_{L}(p,r)+ M f_{L}(p,r)\,,\\
- p\,e^{-A(r)}f_{L}(p,r)&=&\partial_r f_{R}(p,r)+2A^\prime(r)f_{R}(p,r)- M f_{R}(p,r)\,,
\end{eqnarray}
and boundary conditions
\be\label{BCs}
\left.\frac{}{}f_L(p,r)f_R(p,r)\right|_{r_i}=0\,;\,\,i=1,2\,,
\ee
as long as the boundary fields $\psi^0_L$ and $\psi^0_R$ are related by
\be\label{relation}
\slashed{p}\,\psi^0_R(p)=p\frac{f_R(p,r_2)}{f_L(p,r_2)}\psi^0_L(p)\,.
\ee
The boundary conditions arise because, in applying the variational principle, one encounters a total derivative of the form\footnote{One should note that in order to see this term, one should first symmetrise the action Eq.~(\ref{fermionaction}) as in \cite{ConPom}.}
\be
\partial_r(\delta\bar{\psi}_L\gamma^5\psi_R+\delta\bar{\psi}_R\gamma^5\psi_L)\,,
\ee
which must vanish, implying
\be\label{bg5}
\left.\frac{}{}\delta\bar{\psi}_L\gamma^5\psi_R+\delta\bar{\psi}_R\gamma^5\psi_L\right|_{r_i}=0\,.
\ee
Applying Eq.~(\ref{dec}) and Eq.~(\ref{relation}), it is possible to rewrite Eq.~(\ref{bg5}) as
\be
\left.\frac{f_L(p,r)f_R(p,r)}{f_L(p,r_2)f_R(p,r_2)}\delta\bar{\psi}_L^0(p)\gamma^5\psi_R^0(p)+\frac{p f_L(p,r)f_R(p,r)}{\slashed{p}f_L^2(p,r_2)}\delta\bar{\psi}_L^0(p)\gamma^5\psi_L^0(p)\right|_{r_i}=0\nn\,,
\ee
and the second term of this expression vanishes since $\bar{\psi}_L\gamma^5\psi_L=0$. This effectively restores chirality of the zero modes in the boundary theory, which is a property that cannot be defined in five dimensions. This is because Eq.~(\ref{BCs}) forces us to choose Dirichlet boundary conditions for either the left-- or right--handed fields which implies that $f_L=0$ or $f_R=0$ everywhere for the zero modes (note that this is not true for the KK--modes). Working with two fermion fields $\Psi^1$ and $\Psi^2$ and choosing opposite boundary conditions for each field then gives a model where the boundary theory contains massless chiral fields.

\subsection{Massive Light Modes in AdS}

\subsubsection{Approximate Solutions}

In order to give a mass to the zero modes we introduce a boundary term to the fermion action, spontaneously breaking chiral symmetry.
 We choose to add a term in the IR of the form
\be\label{IRBT}
\mathcal{S}_{{\rm IR}}=\int d^4x\int_{r_1}^{r_2}dr\sqrt{-\tilde{g}}\lambda(\bar{\psi}^1_{L}\psi^2_{R}+{\it h.c.})\delta(r-r_1)\,,
\ee
where $\lambda$ has mass dimension $[\lambda]=1$. If $\lambda$ is small, this term can be treated as a perturbation of the chiral model discussed in the previous section and the mass of the light states is
\be\label{mass}
m=e^{4A(r_1)}\lambda N^1_L N^2_R f^{1\,\,0}_L(r_1)f^{2\,\,0}_R(r_1)\,,
\ee
where $f^{i\,\,0}_{L,R}$ is the first term of the Taylor expansion of $f^i_{L,R}$
\be
f^i_{L,R}=f^{i\,\,0}_{L,R}+p f^{i\,\,1}_{L,R}+\cdots\,,
\ee
and describes the zero modes. $N^i_{L,R}$ are the normalization of the zero modes, found by requiring that the 4D kinetic term be canonically normalized. This is constructed by integrating over the extra dimension in the 5D kinetic term
\be\label{gennorm}
\frac{1}{\left(N^i_{L,R}\right)^2}=\int_{r_1}^{r_2}dr\, e^{3A(r)}\left(f^{i\,\,0}_{L,R}\right)^2\,.
\ee
How the light states are localized in the bulk can also be determined by considering the $r$--dependence of the argument of this integral. Note that this definition of the localization includes the warp factor.

\bs

\begin{figure}[t]
\begin{center}
\subfigure{\includegraphics[scale=0.7]{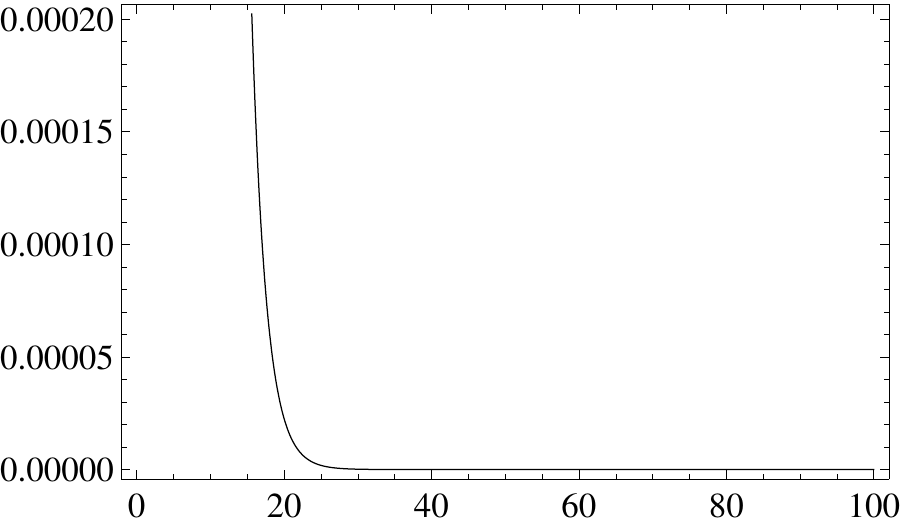}
\Text(-171,115)[]{\small{$e^{3r}(f^0_R)^2$}}
\Text(-15,0)[]{\small{$r$}}}
\subfigure{\includegraphics[scale=0.7]{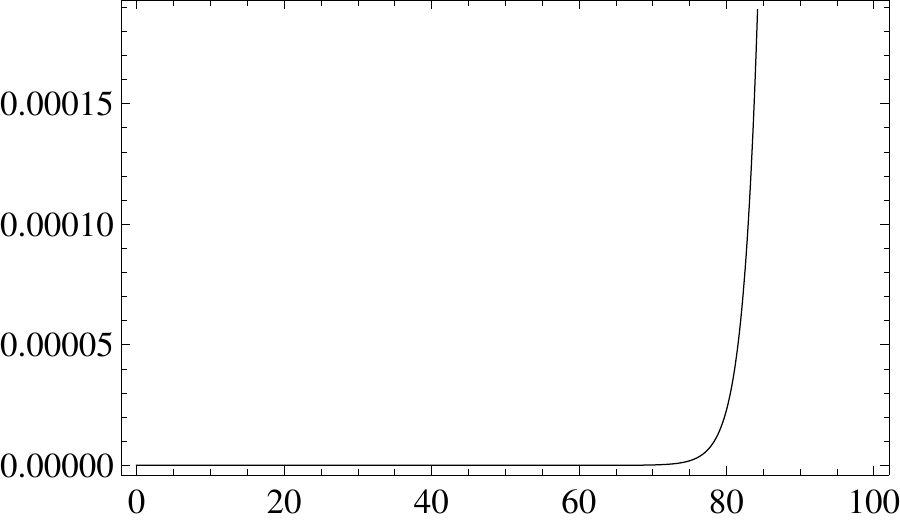}
\Text(-171,115)[]{\small{$e^{3r}(f^0_R)^2$}}
\Text(-15,0)[]{\small{$r$}}}
\subfigure{\includegraphics[scale=0.7]{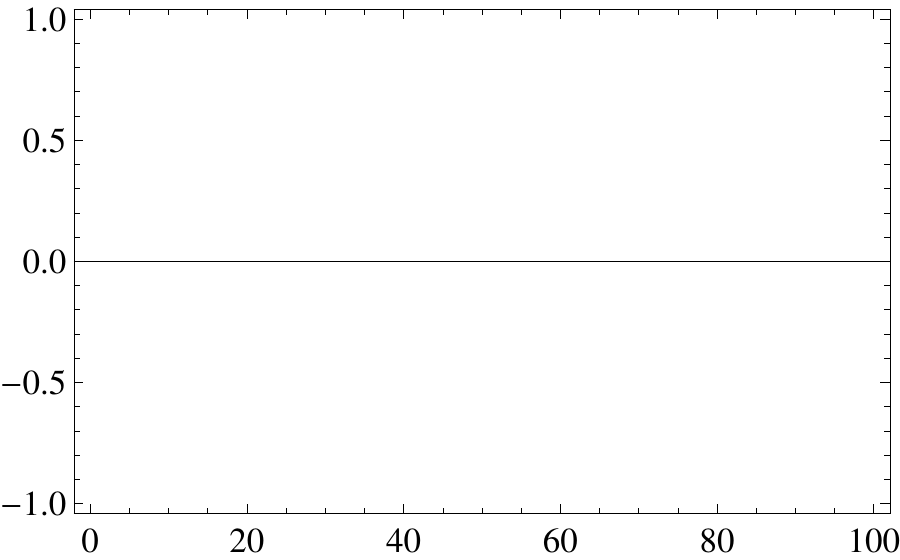}
\Text(-192,115)[]{\small{$e^{3r}(f^0_R)^2$}}
\Text(-15,0)[]{\small{$r$}}}
\end{center}\caption{Localization of a generic right--handed fermion light state for bulk mass $M=1/4$ (first panel), $M=3/4$ (second panel) and $M=1/2$ (third panel), having set $r_2=100$ and $\kappa=1$. Note that left--handed and right--handed fields are related by the transformation $M\rightarrow -M$.} 
\label{locAdS}
\end{figure}

\begin{figure}[t]
\begin{center}
\subfigure{\includegraphics[scale=0.7]{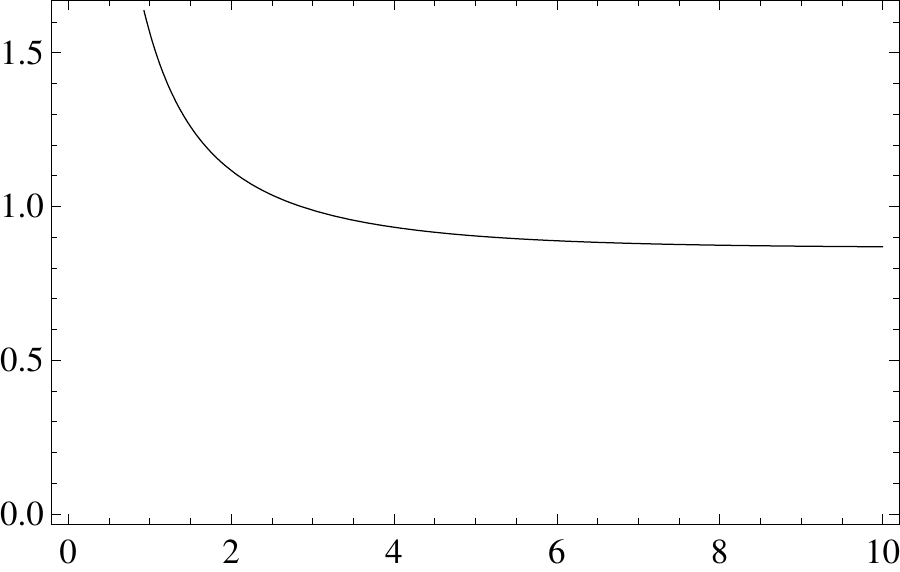}
\Text(-190,110)[]{\small{$m$}}
\Text(-20,0)[]{\small{$r_2$}}}
\subfigure{\includegraphics[scale=0.7]{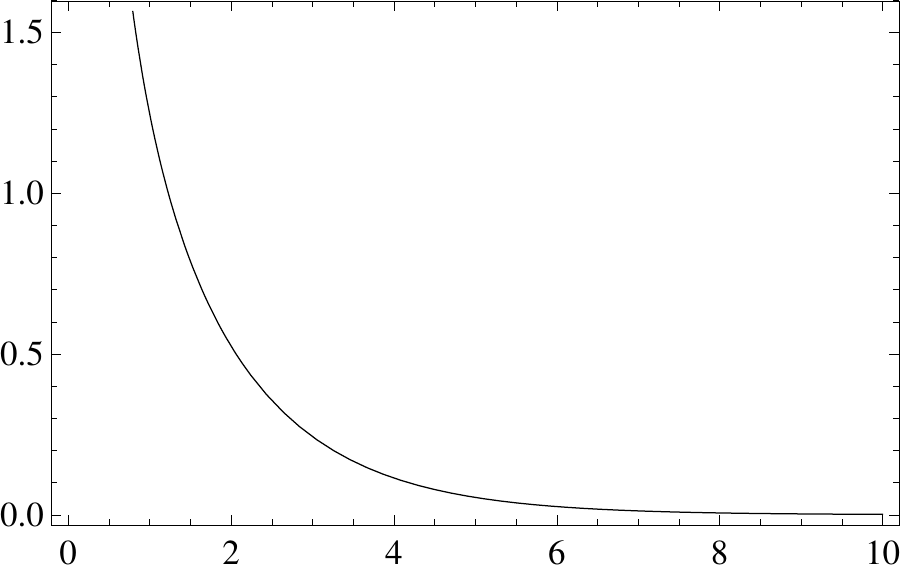}
\Text(10,110)[]{\small{$m$}}
\Text(-20,0)[]{\small{$r_2$}}}
\end{center}\caption{Left panel: plot of the fermion mass against the UV scale $r_2$ for $M_1$ and $M_2$ less than $1/2$. Right panel: plot of the fermion mass against the UV scale $r_2$ for $M_1$ and $M_2$ greater than $1/2$. In both cases we have set all other parameters to one. Note that in the first case the mass tends to a constant as $r_2$ is increased, whereas in the second case it tends to zero.}
\label{approxr2dep}
\end{figure}

\begin{figure}[t]
\begin{center}
\subfigure{\includegraphics[scale=0.7]{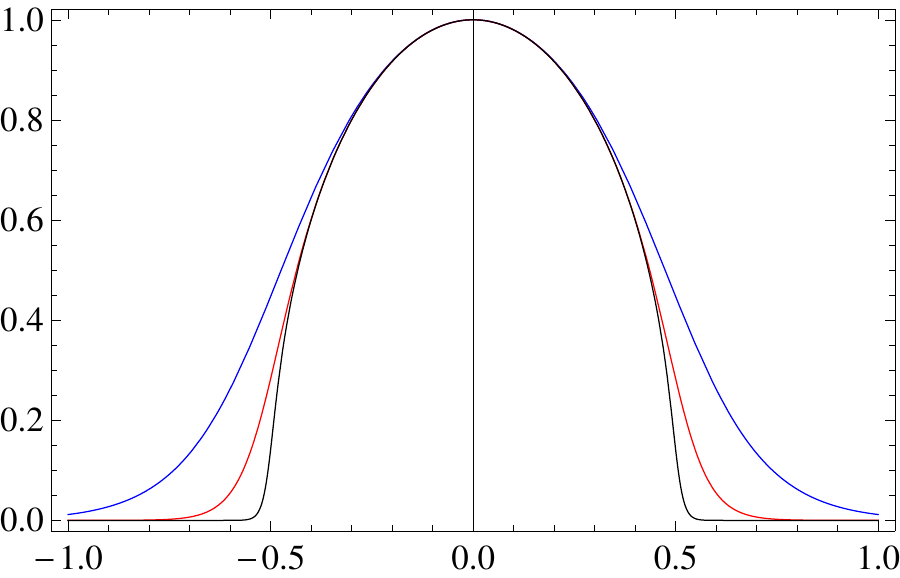}
\Text(-190,110)[]{\small{$m$}}
\Text(-20,0)[]{\small{$M_1$}}}
\subfigure{\includegraphics[scale=0.7]{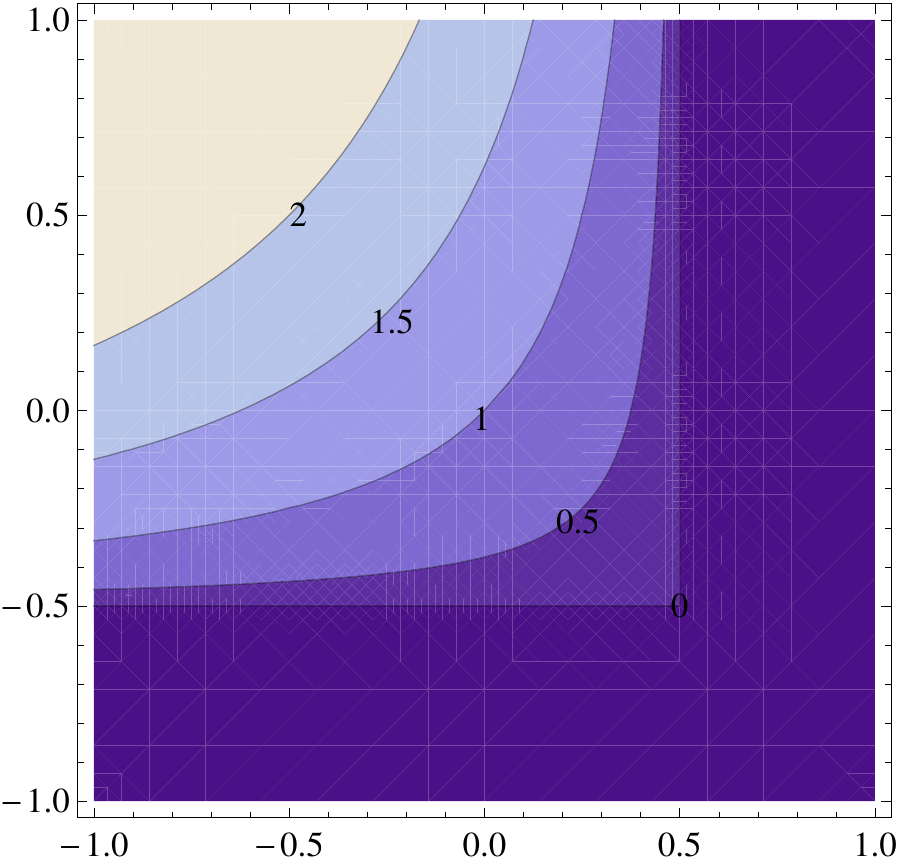}
\Text(10,150)[]{\small{$M_2$}}
\Text(-20,0)[]{\small{$M_1$}}}
\end{center}\caption{Left panel: Plot of the fermion mass against the bulk mass of one of the fields for $M_1=M_2$ and $r_2=10$ (blue curve), $r_2=25$ (red curve) and $r_2=100$ (black curve); with all other parameters set to one. Note that lowering the UV scale increases the fermion mass in the range $|M_1|, |M_2|>1/2$, and that this mass is exponentially suppressed in this region. Right panel: a contour plot of the fermion mass against $M_1$ and $M_2$ for $r_2$ infinite and all other parameters set to one.}
\label{approxmM}
\end{figure}

Working in pure AdS, which is equivalent to setting $\Phi_I=0$ in the deformed background, and choosing the boundary conditions such that
\be
f^{1\,\,0}_L=f^{2\,\,0}_R=0\,,
\ee 
the bulk equations for the zero modes reduce to the decoupled equations
\begin{eqnarray}\label{diffeqn}
\partial_r f^{2\,\,0}_{L}+2\kappa f^{2\,\,0}_{L}+ M_2 f^{2\,\,0}_{L}&=&0\,,\nn\\
\partial_r f^{1\,\,0}_{R}+2\kappa f^{1\,\,0}_{R}- M_1 f^{1\,\,0}_{R}&=&0\,.
\end{eqnarray}
The solutions to these equations are
\begin{eqnarray}
f^{2\,\,0}_L&=&c^2_Le^{-(M_2+2\kappa)r}\,,\nn\\
f^{1\,\,0}_R&=&c^1_Re^{(M_1-2\kappa)r}\,,
\end{eqnarray}
 which are localized as shown in Figure \ref{locAdS} for various choices of the bulk mass $M$ \cite{ConPom}. The normalizations are given by
\begin{eqnarray}\label{norm}
\frac{1}{\left(N^2_L\right)^2}&=&\frac{1-e^{-(2M_2+\kappa)r_2}}{2M_2+\kappa}\,,\nn\\
\frac{1}{\left(N^1_R\right)^2}&=&\frac{e^{(2M_1-\kappa)r_2}-1}{2M_1-\kappa}\,,
\end{eqnarray}
which uniquely determine the integration constants $c^i_{L,R}$. This yields an approximate expression for the mass of the light states, from Eq.~(\ref{mass})
\be
m=\lambda \sqrt{\frac{(2M_2+\kappa)(2M_1-\kappa)}{(1-e^{-(2M_2+\kappa)r_2})(e^{(2M_1-\kappa)r_2}-1)}}\,.
\ee
Note that this expression depends explicitly on the UV scale $r_2$. However, this is not an issue as the expression is well behaved as we take the limit $r_2\rightarrow\infty$. This can be seen in Figure \ref{approxr2dep} which shows that as $r_2$ is taken large, the mass tends to a constant value. If $M_1>1/2$ or $M_2<-1/2$ the fermions will become massless in this limit, which provides a natural mechanism by which the mass can be suppressed. We are also interested in the dependence of the physical mass on the bulk masses $M_1$ and $M_2$: this is presented in Figure \ref{approxmM} for various choices of the UV scale. Of particular interest is the fact that keeping $r_2$ finite allows massive light states for all values of the bulk masses, but the mass is exponentially suppressed for $M_1>1/2$ or $M_2<-1/2$.

\subsubsection{Exact Solution}

\begin{figure}[t]
\begin{center}
\subfigure{\includegraphics[scale=0.7]{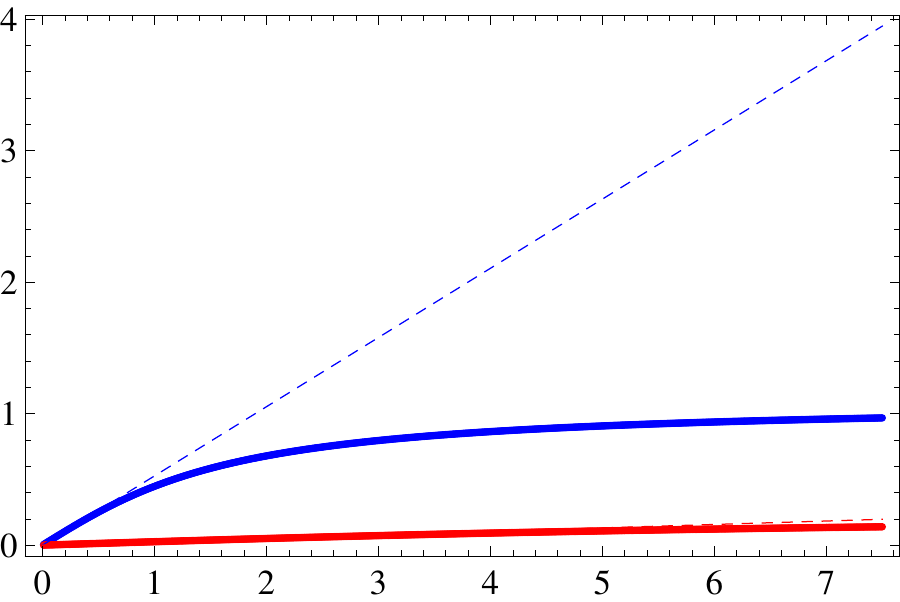}
\Text(-190,110)[]{\small{$m$}}
\Text(-20,0)[]{\small{$\lambda$}}}
\subfigure{\includegraphics[scale=0.7]{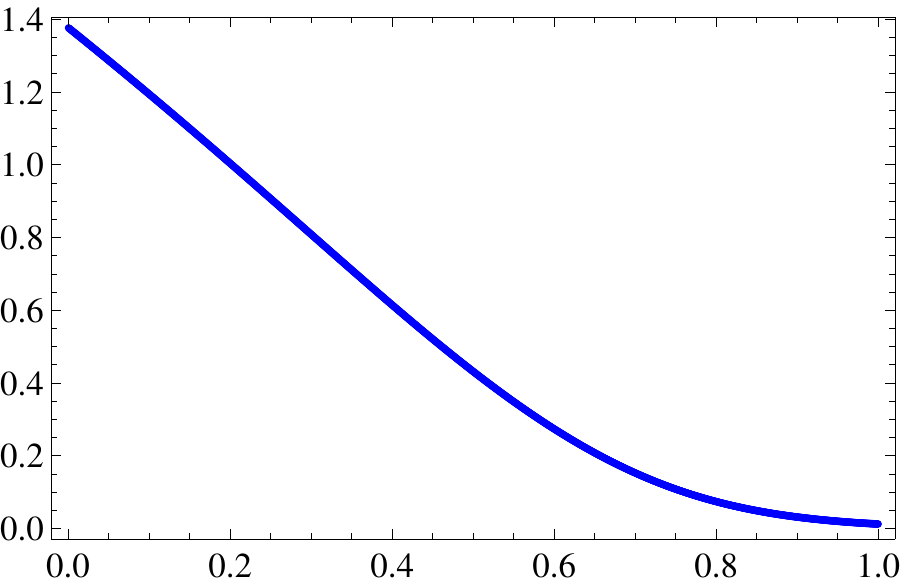}
\Text(10,110)[]{\small{$m$}}
\Text(-20,0)[]{\small{$M$}}}
\end{center}\caption{Left panel: plot of the ground state fermion mass against $\lambda$. The blue curve is for $M=1/4$, the red curve is for $M=3/4$ (where $M_1=-M_2=M$) and the dashed curves are the corresponding approximate solutions. Note that the approximate solutions are good for small values of $\lambda$, but as $\lambda$ is increased further the mass tends to a constant. This cannot be seen from the approximations and gives an upper bound on how much the fermion mass can be increased by increasing $\lambda$.  Right panel: plot of the light fermion mass against the the bulk mass $M$ for $M_1=-M_2=M$ and large $\lambda$ ($\lambda=5$). Note that for $M<1/2$, the mass falls linearly with increasing bulk mass, but for $M>1/2$ the fall off is exponential. The UV scale was taken to be $r_2=6$ and the curvature was set to $\kappa=1$ for both plots.}
\label{fullsolnlightAdS}
\end{figure}

\begin{figure}[t]
\begin{center}
\subfigure{\includegraphics[scale=0.7]{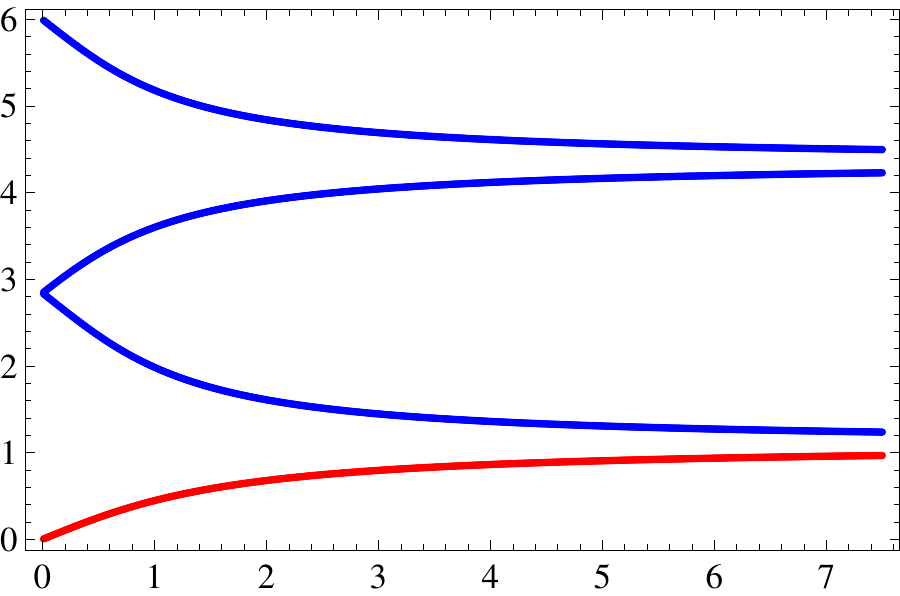}
\Text(-190,110)[]{\small{$m$}}
\Text(-20,0)[]{\small{$\lambda$}}}
\subfigure{\includegraphics[scale=0.7]{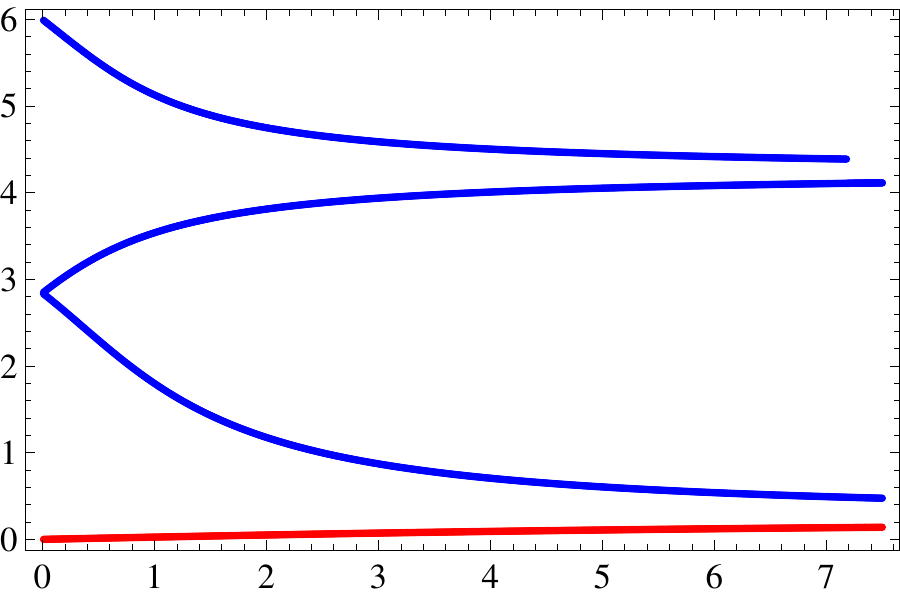}
\Text(10,110)[]{\small{$m$}}
\Text(-20,0)[]{\small{$\lambda$}}}
\end{center}\caption{Left panel: plot of the masses of the first three KK--modes against $\lambda$ for $M_1=-M_2=M=1/4$. The red curve shows the light state for comparison.  Right panel: plot of the masses of the first three KK--modes against $\lambda$ for $M_1=-M_2=M=3/4$. The red curve shows the light state for comparison.}
\label{fullsolnKKAdS}
\end{figure}

The approach of the previous section relies crucially on the the four--dimensional physical mass of the fermion being small. While this simplification may be appealing, it is also prudent to compare with the exact solution. We therefore ask what happens when we include explicitly the IR term in the boundary conditions, without approximation. Working with the fields $\hat{\psi}_{L,R}^i=e^{2A}\psi^i_{L,R}$ simplifies the equations of motion
\begin{eqnarray}
-e^{-A(r)}\slashed{p}\hat{\psi}^i_R+\left(\partial_r+M_i\right)\hat{\psi}^i_L&=&0\,,\nn\\
e^{-A(r)}\slashed{p}\hat{\psi}^i_L+\left(\partial_r-M_i\right)\hat{\psi}^i_R&=&0\,,
\end{eqnarray}
and combining the two equations yields a second order equation for the functions $f^i_{L,R}$
\be\label{order2diff}
\left[1+\frac{e^{2\kappa r}}{p^2}\left(\partial^2_r+\kappa \partial_r\pm M_i\kappa-(M_i)^2\right)\right]\hat{f}^i_{L,R}(p,r)=0\,,
\ee
which has general solutions of the form
\begin{eqnarray}
\hat{f}^i_L&=&\sqrt{p}e^{-\frac{\kappa r}{2}}\left(a^i_L J_{\frac{M_i}{\kappa}-\frac{1}{2}}\left(\frac{e^{-\kappa r}p}{\kappa}\right)-b^i_L Y_{\frac{M_i}{\kappa}-\frac{1}{2}}\left(\frac{e^{-\kappa r}p}{\kappa}\right)\right)\,,\nn\\
\hat{f}^i_R&=&\sqrt{p}e^{-\frac{\kappa r}{2}}\left(a^i_R J_{-\frac{M_i}{\kappa}-\frac{1}{2}}\left(\frac{e^{-\kappa r}p}{\kappa}\right)-b^i_R Y_{-\frac{M_i}{\kappa}-\frac{1}{2}}\left(\frac{e^{-\kappa r}p}{\kappa}\right)\right)\,.
\end{eqnarray}
IR boundary terms of the form of Eq.~(\ref{IRBT}) have no effect on the UV boundary conditions
\begin{eqnarray}\label{uvbc}
\left.\frac{}{}\hat{f}^1_L\right|_{r_2}&=&0\,,\nn\\
\left.\frac{}{}\hat{f}^2_R\right|_{r_2}&=&0\,,\nn\\
\left.\frac{}{}\left(\partial_r+M_2\right)\hat{f}^2_L\right|_{r_2}&=&0\,,\nn\\
\left.\frac{}{}\left(\partial_r-M_1\right)\hat{f}^1_R\right|_{r_2}&=&0\,,
\end{eqnarray}
where the two additional boundary conditions come from requiring that the bulk equation be satisfied on the UV boundary. The IR boundary conditions can be found using the variational principle\footnote{A more careful treatment of the boundary conditions is presented in \cite{csaki et al}. It should be noted that, while this approach is much simpler, it yields the same result.}
\begin{eqnarray}\label{irbc}
\left.\frac{}{}\hat{f}^1_L-\lambda\hat{f}^2_L\right|_{r_1}&=&0\,,\nn\\
\left.\frac{}{}\hat{f}^2_R+\lambda\hat{f}^1_R\right|_{r_1}&=&0\,,\nn\\
\left.\frac{}{}-e^{-A(r)}p\hat{f}^2_R+\left(\partial_r+M_2\right)\hat{f}^2_L\right|_{r_1}&=&0\,,\nn\\
\left.\frac{}{}e^{-A(r)}p\hat{f}^1_L+\left(\partial_r-M_1\right)\hat{f}^1_R\right|_{r_1}&=&0\,,
\end{eqnarray}
where, again, the last two boundary conditions are found by requiring that the bulk equation be satisfied at the IR boundary\footnote{One should note that imposing the first two boundary conditions of Eq.~(\ref{uvbc}) on the bulk equation at the UV boundary yields the second two. This is why the IR and UV boundary conditions look different.}. This leaves us with eight unknown integration constants (the $a^i_{L,R}$ and $b^i_{L,R}$) and eight constraints on the system. Since the equations of motion are linear, we can always normalize such that one of these integration constants is one, meaning the system is over constrained. We can, therefore, use one of these boundary conditions to extract the spectrum of states present in our model. As an illustration, we consider the case $M_1=-M_2=M$ and set the curvature $\kappa=1$. Figure \ref{fullsolnlightAdS} shows the dependence of the light state on $\lambda$ and $M$, while Figure \ref{fullsolnKKAdS} shows the KK--modes as a function of $\lambda$ for various choices of the other parameters. Note that the approximate solutions are indeed satisfactory as long as $\lambda$ is small. However, when $\lambda$ is large the mass of the light state tends to a constant value, placing an upper bound on how much the mass can be increased by dialing $\lambda$. This has important consequences for the phenomenology of the top quark (for a discussion see \cite{nomburd}).

\subsection{Fermions in the Deformed Background}

\subsubsection{Approximate Solutions}

\begin{figure}[t]
\begin{center}
\subfigure{\includegraphics[scale=0.7]{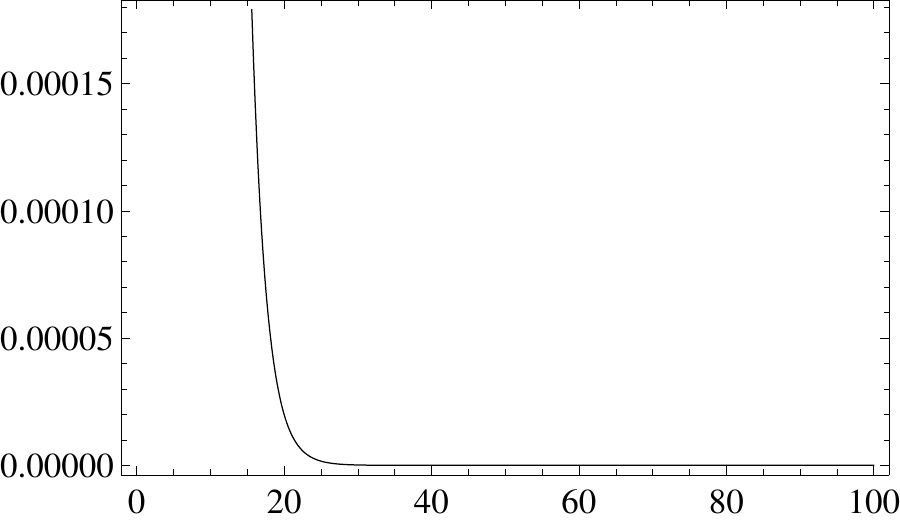}
\Text(-171,115)[]{\small{$e^{3r}(f^0_R)^2$}}
\Text(-15,0)[]{\small{$r$}}}
\subfigure{\includegraphics[scale=0.7]{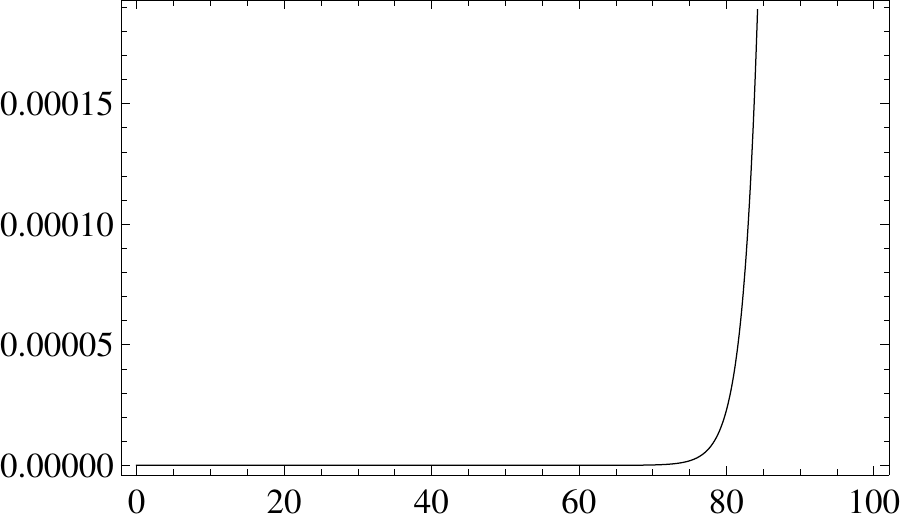}
\Text(-171,115)[]{\small{$e^{3r}(f^0_R)^2$}}
\Text(-15,0)[]{\small{$r$}}}
\subfigure{\includegraphics[scale=0.7]{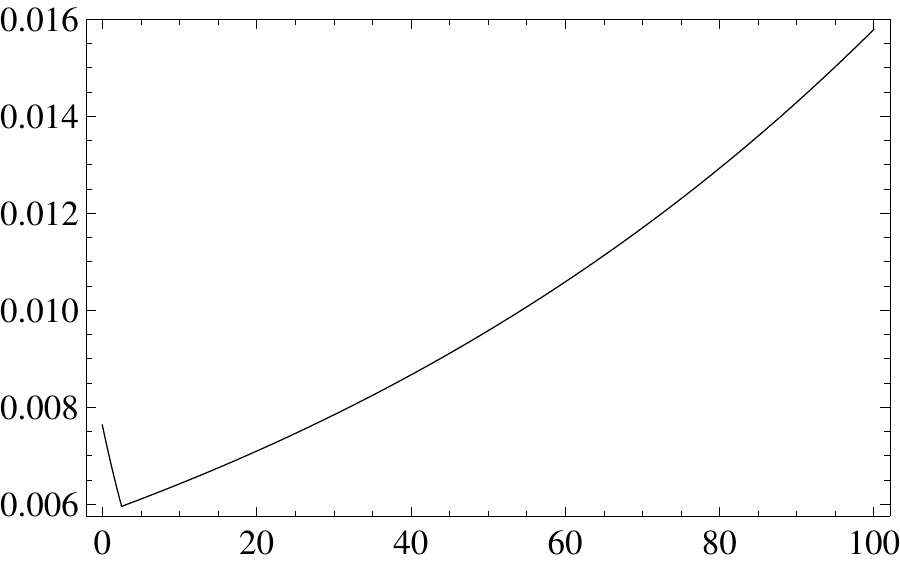}
\Text(-197,115)[]{\small{$e^{3r}(f^0_R)^2$}}
\Text(-15,0)[]{\small{$r$}}}
\end{center}\caption{Localization of a generic right--handed fermion light state in the deformed background for bulk mass $M=1/4$ (first panel), $M=3/4$ (second panel) and $M=0.505$ (third panel), having set $r_2=100$, $\Delta=1$, $\Phi_I=1$, $r_*=2.5$ and $\kappa_1=1$. Note that as well as being localized in the IR or UV, it is also possible to have intermediate solutions where the fermion wavefunction is peaked at both boundaries.} 
\label{1sloc}
\end{figure}

\begin{figure}[t]
\begin{center}
\subfigure{\includegraphics[scale=0.7]{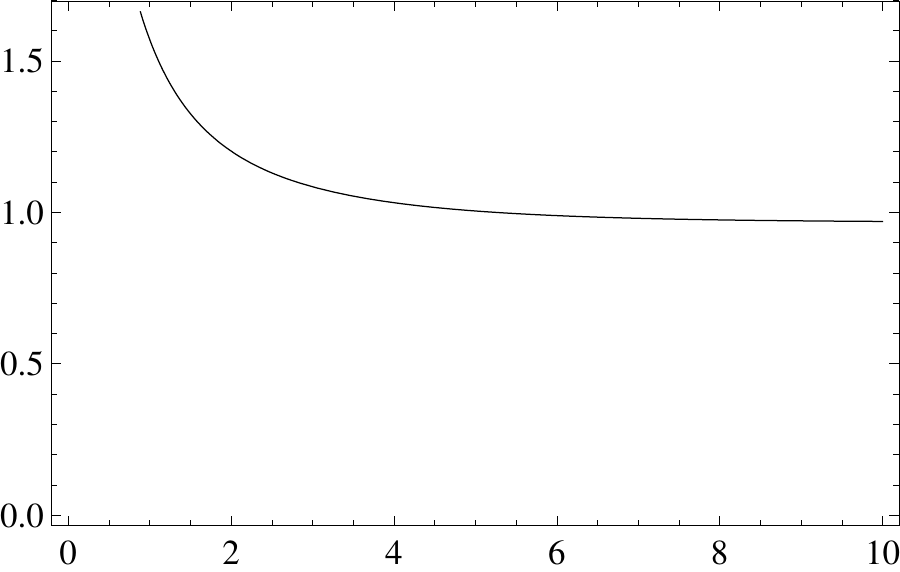}
\Text(-190,110)[]{\small{$m$}}
\Text(-20,0)[]{\small{$r_2$}}}
\subfigure{\includegraphics[scale=0.7]{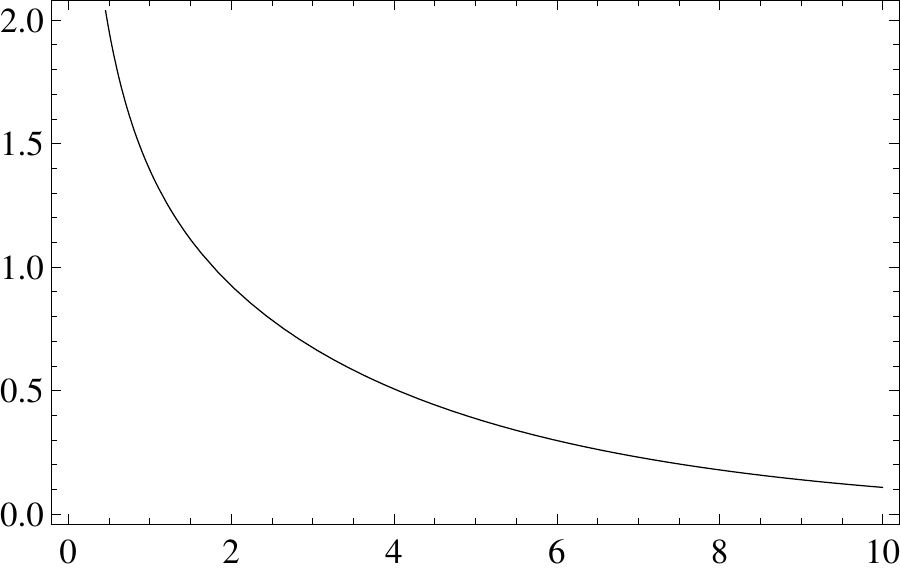}
\Text(10,110)[]{\small{$m$}}
\Text(-20,0)[]{\small{$r_2$}}}
\end{center}\caption{Left panel: plot of the fermion mass against the UV scale $r_2$ in the deformed background for $M_1$ and $M_2$ less than $1/2$. Right panel: plot of the fermion mass against the UV scale $r_2$ for $M_1$ and $M_2$ greater than $1/2$. For both plots we use $\lambda=1$, $\kappa_1=1$, $\Delta=1$, $\Phi_I=1$ and $r_*=2.5$ and choose $M_1=M_2=1/4$ for the first plot and $M_1=M_2=3/4$ for the second. Note that in the first case the mass tends to a constant as $r_2$ is increased, whereas in the second case it tends to zero. This is qualitatively the same as the ADS case.}
\label{1sapproxr2}
\end{figure}

\begin{figure}[t]
\begin{center}
\subfigure{\includegraphics[scale=0.7]{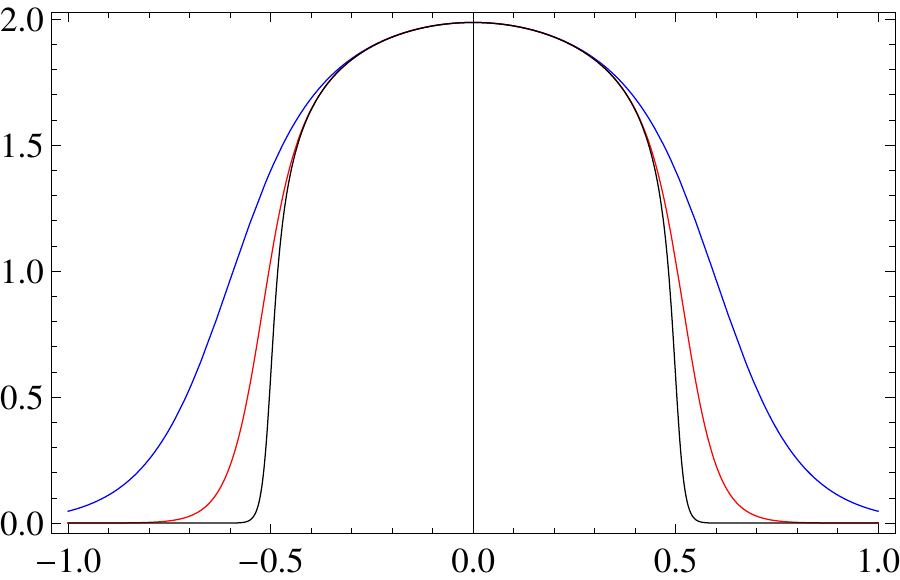}
\Text(-190,110)[]{\small{$m$}}
\Text(-20,0)[]{\small{$M_1$}}}
\subfigure{\includegraphics[scale=0.7]{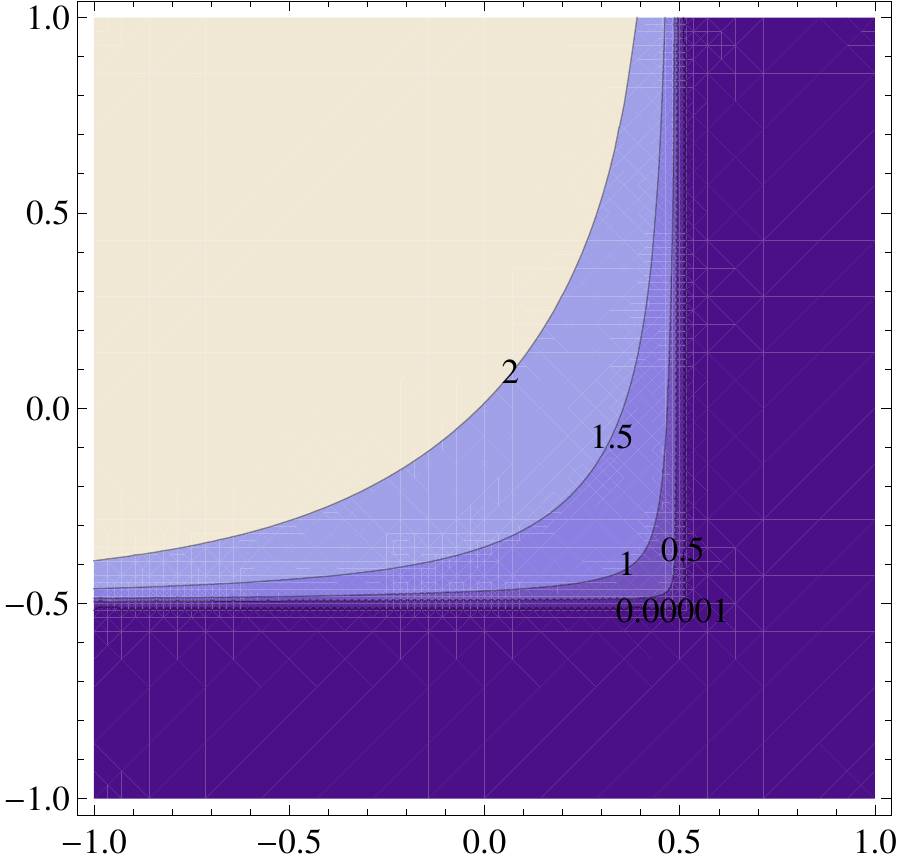}
\Text(10,150)[]{\small{$M_2$}}
\Text(-20,0)[]{\small{$M_1$}}}
\end{center}\caption{Left panel: plot of the fermion mass against the bulk mass of one of the fields in the deformed background for $M_1=M_2$ and $r_2=10$ (blue curve), $r_2=25$ (red curve) and $r_2=100$ (black curve). Note that lowering the UV scale increases the fermion mass in the range $|M_1|, |M_2|>1/2$, and that this mass is exponentially suppressed in this region. Right panel: a contour plot of the fermion mass against $M_1$ and $M_2$ for $r_2\rightarrow\infty$. In both plots the other parameters are set to $\lambda=1$, $\kappa_1=1$, $\Delta=3$, $\Phi_I=\sqrt{3}$, $r_*=2.5$.}
\label{1sapproxM}
\end{figure}

We now turn our attention to fermions propagating in a background deformed by the backreaction upon the metric of a scalar field as in Eq.~(\ref{phicl}). This background is taken from the model explored in \cite{RL and MP} and outlined in section 2. Proceeding as in the AdS case, we find that the light state wavefunctions are approximately
\begin{eqnarray}
f^{2\,\,0}_L&=&\left\{\begin{array}{l} c_1e^{-(2\kappa_0+M_2)r}\,,\,\,r<r_*\\
c_1e^{-2(\kappa_0-\kappa_1)r_*}e^{-(2\kappa_1+M_2)r}\,,\,\,r>r_*\\
\end{array}\right.\,,\\
f^{1\,\,0}_R&=&\left\{\begin{array}{l} c_2e^{(M_1-2\kappa_0)r}\,,\,\,r<r_*\\
c_2e^{-2(\kappa_0-\kappa_1)r_*}e^{(M_1-2\kappa_1)r}\,,\,\,r>r_*\\
\end{array}\right.\,.
\end{eqnarray}
The fermions in this case can be localized in the bulk as shown in Figure \ref{1sloc}. Each panel of Figure \ref{1sloc} is generated by fixing the parameters $r_2$, $\Delta$, $\Phi_I$, $r_*$ and $\kappa_1$ and plotting the dependence of the argument of the normalization integral (for the right--handed zero mode) for various choices of the bulk mass. These choices are such that the bulk mass lies in one of the intervals $M<\kappa_1/2$, $\kappa_1/2<M<\kappa_0/2$ or $M>\kappa_0/2$, where $\kappa_0$ is determined by $\Delta$, $\Phi_I$ and $r_*$. The mass is given by
\begin{eqnarray}\label{mass1s}
m&=&\lambda \left(\frac{1-e^{-\left(\kappa _0+2 M_2\right) r_*}}{\kappa _0+2
   M_2}+\frac{e^{\left(\kappa _1-\kappa _0\right) r_*} \left(e^{-\left(\kappa _1+2
   M_2\right) r_*}-e^{- \left(\kappa _1+2 M_2\right)r_2}\right)}{\kappa
   _1+2 M_2}\right)^{-\frac{1}{2}}\nn\\&&  \times\left(\frac{1-e^{\left(2 M_1-\kappa _0\right) r_*}}{\kappa _0-2 M_1}+\frac{e^{\left(\kappa _1-\kappa
   _0\right) r_*} \left(e^{\left(2 M_1 -\kappa _1\right)
   r_*}-e^{\left(2 M_1 -\kappa _1\right) r_2}\right)}{\kappa _1-2 M_1}\right)^{-\frac{1}{2}}
\end{eqnarray}
As before the expression is well behaved in the limit $r_2\rightarrow\infty$ but taking this limit still yields massless fermions for $M_1>1/2$ or $M_2<-1/2$. Figures \ref{1sapproxr2} and \ref{1sapproxM} show the dependence of the mass on the UV scale and the bulk masses respectively for $\lambda$, $\kappa_1$, $\Delta$, $\Phi_I$ and $r_*$ fixed and $M_1=M_2$. In particular one should note that the results are qualitatively similar to the AdS results and deforming the background in this manner, for the choice of parameter $\Delta=3,\,\Phi_I=\sqrt{3}$ and $r_*=2.5$, results in a mild enhancement of the mass of the light states.


\subsubsection{Numerical Solution}

\begin{figure}[t]
\begin{center}
\subfigure{\includegraphics[scale=0.7]{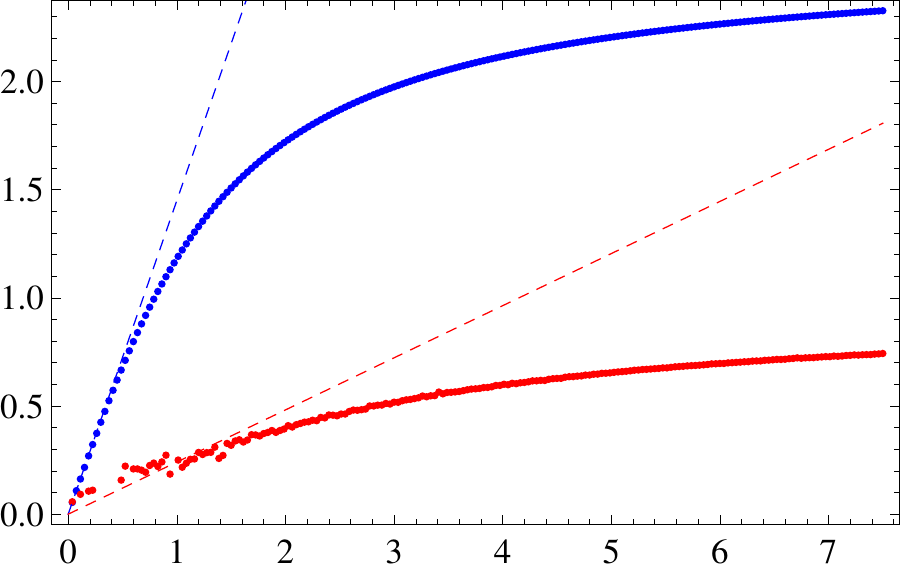}
\Text(-190,110)[]{\small{$m$}}
\Text(-20,0)[]{\small{$\lambda$}}}
\subfigure{\includegraphics[scale=0.7]{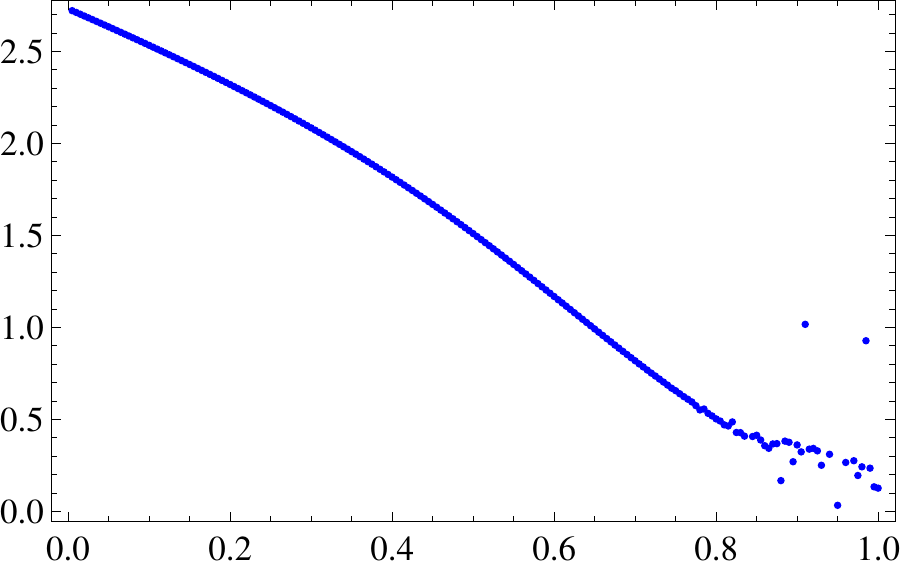}
\Text(10,110)[]{\small{$m$}}
\Text(-20,0)[]{\small{$M$}}}
\end{center}\caption{Left panel: plot of the ground state fermion mass against $\lambda$ in the deformed background. The blue curve is for $M=1/4$, the red curve is for $M=3/4$ (where $M_1=-M_2=M$) and the dashed curves are the corresponding approximate solutions. Note that the approximate solutions are good for small values of $\lambda$, but as $\lambda$ is increased further the mass tends to a constant. This cannot be seen from the approximations and gives an upper bound on how much the fermion mass can be increased by increasing $\lambda$.  Right panel: plot of the light fermion mass against the the bulk mass $M$ in the deformed background for $M_1=-M_2=M$ and large $\lambda$ ($\lambda=5$). Note that for $M<1/2$, the mass falls linearly with increasing bulk mass, but for $M>1/2$ the fall off is exponential. The UV scale was taken to be $r_2=6$ and the curvature was set to $\kappa_1=1$, $\Delta=3$, $\Phi_I=\sqrt{3}$ and $r_*=2.5$ for both plots. Both plots show an enhancement over the AdS results.}
\label{1slight}
\end{figure}

\begin{figure}[t]
\begin{center}
\subfigure{\includegraphics[scale=0.7]{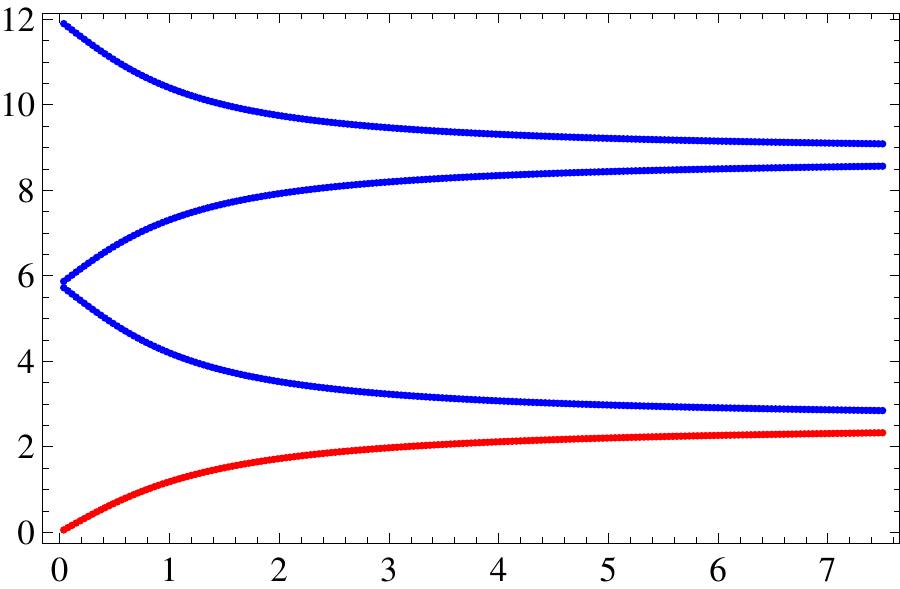}
\Text(-190,110)[]{\small{$m$}}
\Text(-20,0)[]{\small{$\lambda$}}}
\subfigure{\includegraphics[scale=0.7]{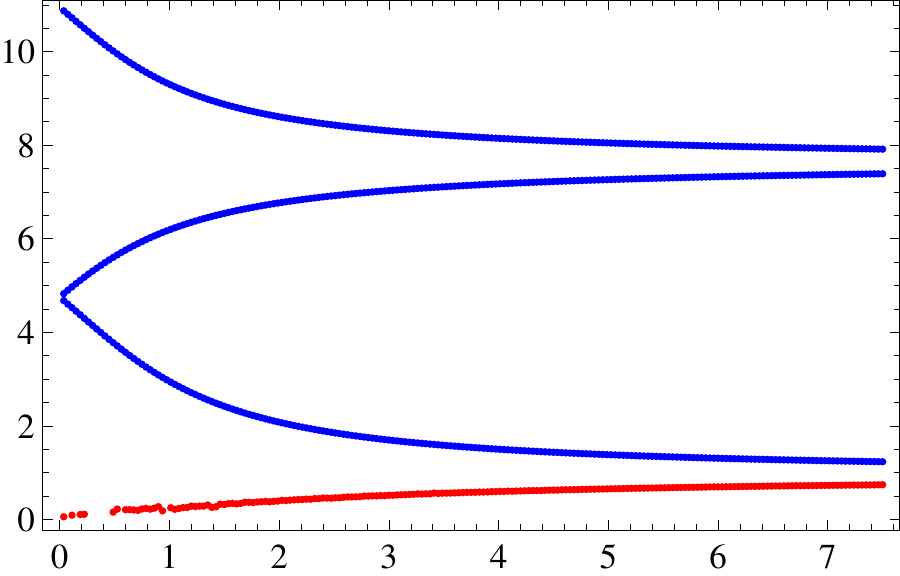}
\Text(10,110)[]{\small{$m$}}
\Text(-20,0)[]{\small{$\lambda$}}}
\end{center}\caption{Left panel: plot of the masses of the first three KK--modes against $\lambda$ for $M_1=-M_2=M=1/4$ in the deformed background. The red curve shows the light state for comparison.  Right panel: plot of the masses of the first three KK--modes against $\lambda$ for $M_1=-M_2=M=3/4$ in the deformed background. The red curve shows the light state for comparison. The UV scale was taken to be $r_2=6$ and the curvature was set to $\kappa_1=1$, $\Delta=3$, $\Phi_I=\sqrt{3}$ and $r_*=2.5$ for both plots. Both plots show a mild enhancement over the AdS results.}
\label{1skk}
\end{figure}

In order to find solutions in the deformed background which include the IR term in the boundary conditions we treat solutions in the regions $r<r_*$ and $r>r_*$ separately and match the solutions at $r=r_*$. In order to do this we need to modify Eq.~(\ref{order2diff}) by replacing $\kappa$ with $\kappa_0$ for $r<r_*$ and by replacing $\kappa$ with $\kappa_1$ and $r$ with $r+(\kappa_0-\kappa_1)r_*$ for $r>r_*$. The general solutions in the two regions are
\begin{eqnarray}
\hat{f}^i_L&=&\sqrt{p}e^{-\frac{\kappa_0 r}{2}}\left(a^i_L J_{\frac{M_i}{\kappa_0}-\frac{1}{2}}\left(\frac{e^{-\kappa_0 r}p}{\kappa_0}\right)-b^i_L Y_{\frac{M_i}{\kappa_0}-\frac{1}{2}}\left(\frac{e^{-\kappa_0 r}p}{\kappa_0}\right)\right)\,,\nn\\
\hat{f}^i_R&=&\sqrt{p}e^{-\frac{\kappa_0 r}{2}}\left(a^i_R J_{-\frac{M_i}{\kappa_0}-\frac{1}{2}}\left(\frac{e^{-\kappa_0 r}p}{\kappa_0}\right)-b^i_R Y_{-\frac{M_i}{\kappa_0}-\frac{1}{2}}\left(\frac{e^{-\kappa_0 r}p}{\kappa}\right)\right)\,,
\end{eqnarray}
when $r<r_*$, and
\begin{eqnarray}
\hat{f}^i_L&=&\sqrt{p}e^{-\frac{\kappa_1 r+(\kappa_0-\kappa_1)r_*}{2}}\left(c^i_L J_{\frac{M_i}{\kappa_1}-\frac{1}{2}}\left(\frac{e^{-\kappa_1 r-(\kappa_0-\kappa_1)r_*}p}{\kappa_1}\right)\right.\nn\\
&&\left.-d^i_L Y_{\frac{M_i}{\kappa_1}-\frac{1}{2}}\left(\frac{e^{-\kappa_1 r-(\kappa_0-\kappa_1)r_*}p}{\kappa_1}\right)\right)\,,\nn\\
\hat{f}^i_R&=&\sqrt{p}e^{-\frac{\kappa_1 r+(\kappa_0-\kappa_1)r_*}{2}}\left(c^i_R J_{-\frac{M_i}{\kappa_1}-\frac{1}{2}}\left(\frac{e^{-\kappa_1 r-(\kappa_0-\kappa_1)r_*}p}{\kappa_1}\right)\right.\nn\\
&&\left.-d^i_R Y_{-\frac{M_i}{\kappa_1}-\frac{1}{2}}\left(\frac{e^{-\kappa_1 r-(\kappa_0-\kappa_1)r_*}p}{\kappa_1}\right)\right)\,,
\end{eqnarray}
when $r>r_*$. Having found these general solutions, we now compute the integration constants $a_{L,R}^i$, $b_{L,R}^i$, $c_{L,R}^i$ and $d_{L,R}^i$ by applying the boundary conditions Eq.~(\ref{uvbc}) and Eq.~(\ref{irbc}), as well as additional conditions found by matching the solutions, and their derivatives, at $r=r_*$. We again focus on the situation $M_1=-M_2=M$ and present the dependence of the mass of the light states on $\lambda$ and $M$ in Figure \ref{1slight}, and the dependence of the mass of the KK--modes on $\lambda$ in Figure \ref{1skk}. In both these plots we fix the UV scale $r_2$ and the parameters responsible for determining the curvature $\kappa_1$, $\Delta$, $\Phi_I$ and $r_*$. We note that, as for the approximate solutions, the results are qualitatively similar to the AdS case, and the increase in curvature in the IR causes an enhancement of the fermion mass spectrum.

\subsection{The Role of the IR Curvature}

\begin{figure}[ht!]
\begin{center}
\subfigure{\includegraphics[scale=0.7]{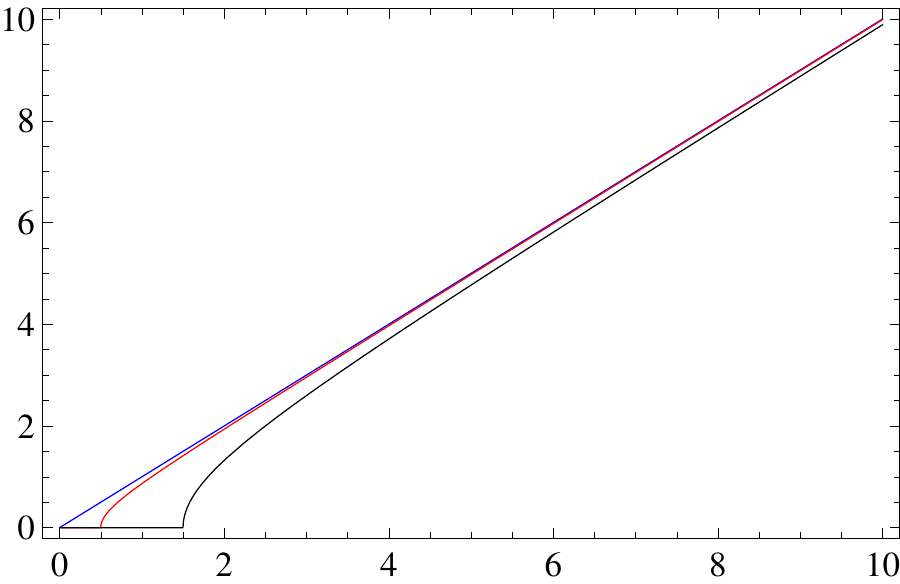}
\Text(-190,110)[]{\small{$m$}}
\Text(-20,0)[]{\small{$\kappa$}}}
\subfigure{\includegraphics[scale=0.7]{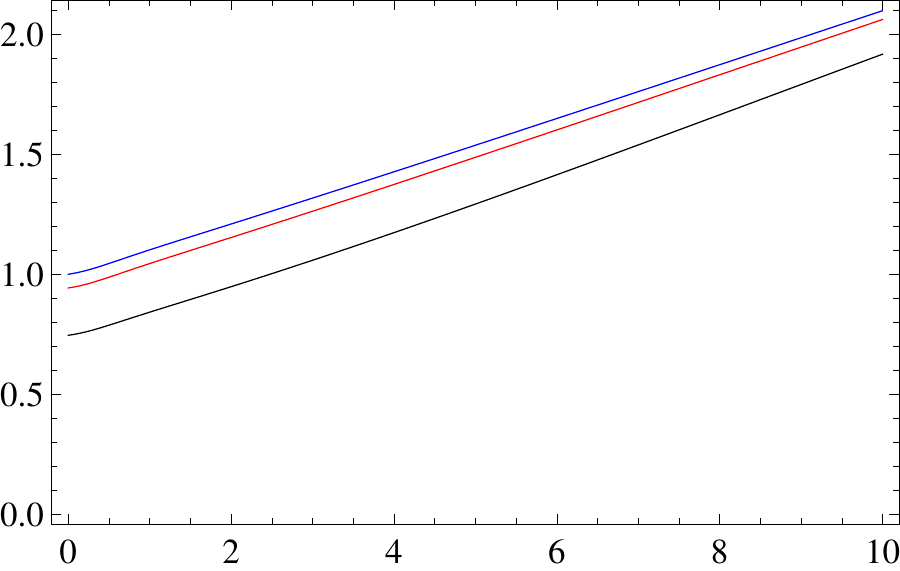}
\Text(10,110)[]{\small{$m$}}
\Text(-20,0)[]{\small{$\Delta$}}}
\subfigure{\includegraphics[scale=0.7]{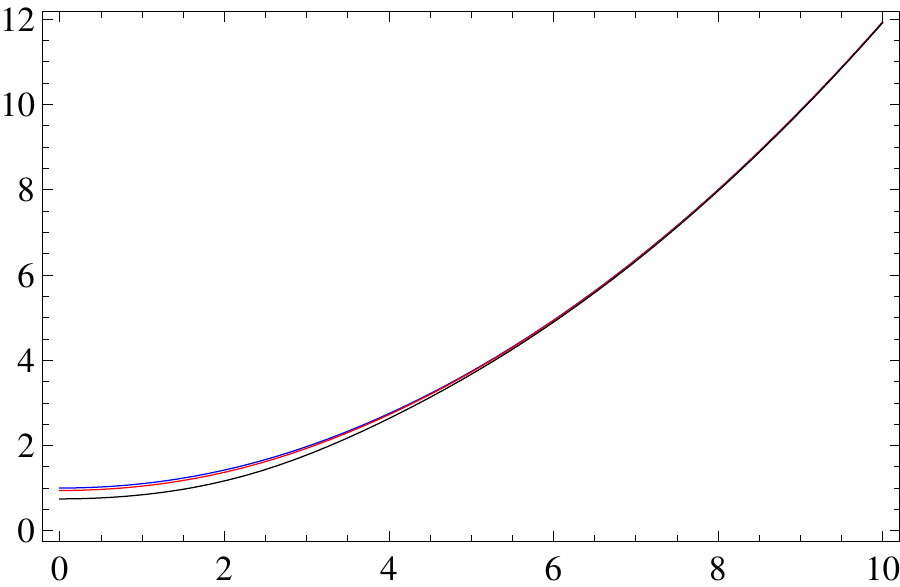}
\Text(-190,110)[]{\small{$m$}}
\Text(-20,0)[]{\small{$\Phi_I$}}}
\subfigure{\includegraphics[scale=0.7]{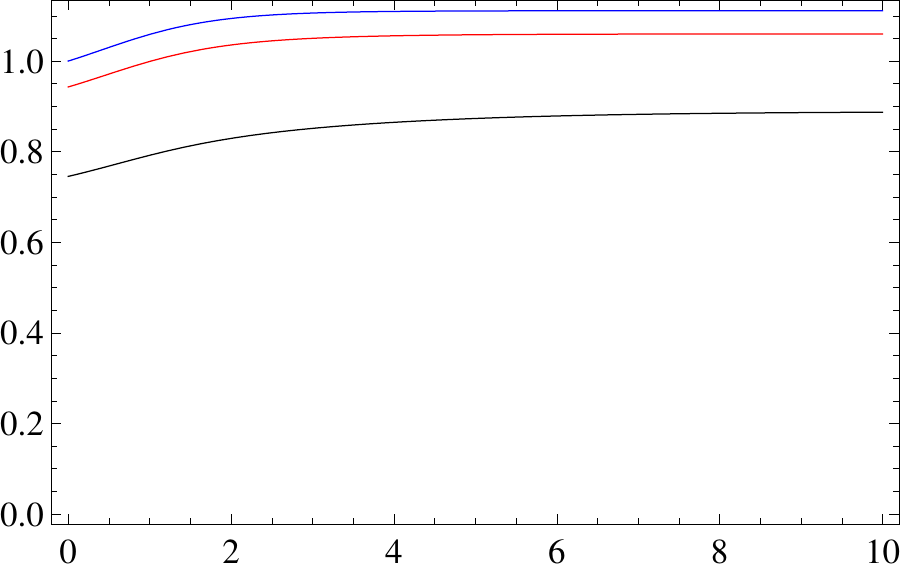}
\Text(10,110)[]{\small{$m$}}
\Text(-20,0)[]{\small{$r_*$}}}
\end{center}\caption{First panel: plot of the dependence of the mass on $\kappa$ in the AdS case for $r_2\rightarrow\infty$. The blue curve is for $M_1=M_2=0$, the red curve is for $M_1=M_2=1/4$ and the black curve is for $M_1=M_2=3/4$. Notice that the derivative of $m(\kappa)$ contains a discontinuity at the point $\kappa=2M$.  As the curvature is increased beyond this point, fermions that were originally massless gain a mass. This is because the limit $r_2\rightarrow\infty$ causes the four--dimensional mass to tend to zero for $M/\kappa>1/2$, while the four--dimensional mass tends to a finite constant if $M/\kappa<1/2$. In essence, $\kappa$ controls the size of the window of bulk masses for which fermions are massive. Second panel: plot of the dependence of the mass on $\Delta$ in the deformed background with $r_2\rightarrow\infty$. Third panel: plot of the dependence of the mass on $\Phi_I$ in the deformed background with $r_2\rightarrow\infty$. Fourth panel: plot of the dependence of the mass on $r_*$ in the deformed background with $r_2\rightarrow\infty$. In the last three plots the blue curve is for $M_1=M_2=0$, the red curve is for $M_1=M_2=1/6$ and the black curve is for $M_1=M_2=1/3$, and we choose the remaining parameters concerning the curvature from the set $\{\kappa_1=1, \Delta=1, \Phi_I=1, r_*=2.5\}$}
\label{curv}
\end{figure}

In the previous sections we have assumed that $\kappa=1$ for AdS and $\kappa_1=1$, $\Delta=3$, $\Phi_I=\sqrt{3}$ and $r_*=2.5$ for the deformed background, where $\Delta$, $\Phi_I$ and $r_*$ determine $\kappa_0$. This choice of parameters has been used to calculate the dependence of the fermion masses on $\lambda$ and the bulk masses of the left and right--handed fields so that comparisons can be made between approximate and exact results and between AdS and the deformed background. Since the role of these parameters has been largely ignored thus far, it is prudent to ask now what effect they have. To simplify the discussion we begin by taking the limit $r_2\rightarrow\infty$ and consider approximate solutions. In this case $\kappa$ (or alternatively $\kappa_1$) controls the range of the bulk mass for which the fermions are massive. This is because the parameter controlling the bulk dynamics of the fermions is not really the bulk mass $M$, but rather $M/\kappa$ (or $M/\kappa_1$). The fermion mass is also approximately linearly dependent on $\kappa$ ($\kappa_1$), as can be seen from the first panel of Figure \ref{curv}. Also of interest is the fact that by considering the deformed background, which introduces a region $r<r_*$ of AdS with increased curvature, we saw an enhancement of the fermion masses. The curvature in this region is controlled by the parameters $\Delta$, $\Phi_I$ and $r_*$ and as such, we also include plots the dependence of the fermion masses on these parameters in this scheme in Figure \ref{curv}.

\bs

To understand this effect, it is first necessary to understand what effect the IR curvature has on the localization of the fermions. In the AdS case, the picture is quite simple: if $M_L<-1/2$ ($M_R>1/2$) the left-- (right--)handed fermion is localized in the UV. For $M_L>-1/2$ ($M_R<1/2$) it is localized in the IR. While $M_L=-1/2$ ($M_R=1/2$) is the classical solution. In this case the fermion profile is flat. Turning on additional sources of IR curvature alters this picture. The fermion is still localized in the IR for $M_L>-1/2$ ($M_R<1/2$), but is now localized in the UV for $M_L<-\kappa_0/2$ ($M_R>\kappa_0/2$). This introduces a new intermediate region where the fermion is localized in the UV but a peak forms at the IR boundary. It is this effect that is responsible for the enhancement of the fermion mass spectrum, since the mass is controlled by the overlap of the left-- and right--handed fermions and the VEV living on the IR boundary.
 
\section{Coupling to $Z$}

\begin{figure}[t]
\begin{center}
\subfigure{\includegraphics[scale=0.7]{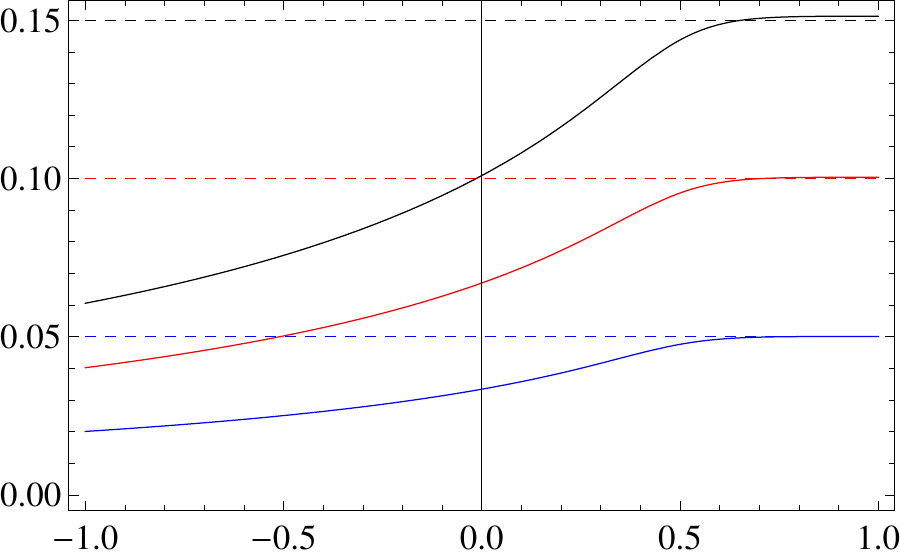}
\Text(-190,110)[]{\small{$I$}}
\Text(-20,0)[]{\small{$M_{R}$}}}
\subfigure{\includegraphics[scale=0.7]{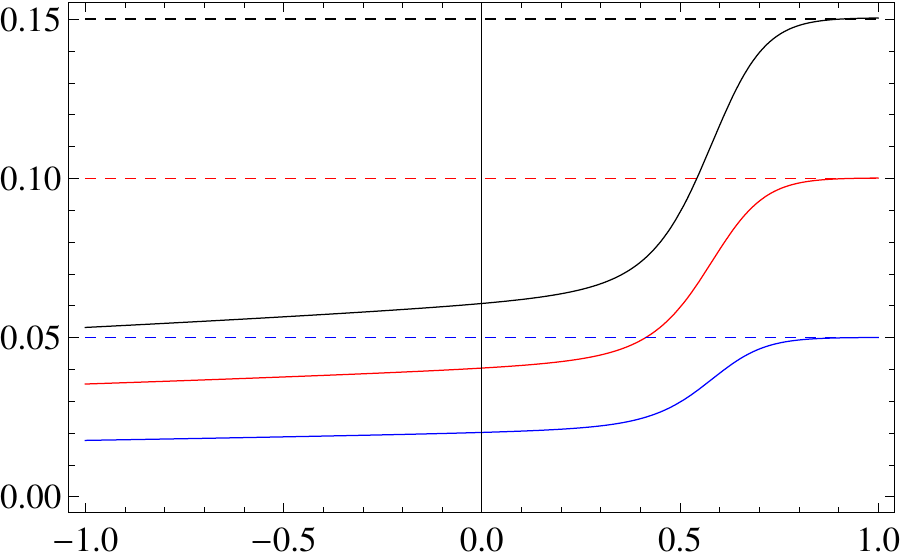}
\Text(10,110)[]{\small{$I$}}
\Text(-20,0)[]{\small{$M_{R}$}}}
\end{center}\caption{Plots of the integral in Eq.~(\ref{int}) against the bulk mass for the left and right--handed fermions, which controls the gauge coupling, in the limit of large $\Omega$. The first plot is for $\delta\kappa=0$, while the second plot is for $\delta\kappa=2$.In both cases we set $r_2=10$ and $r_*=2.5$. The blue curves correspond to $\varepsilon=0.05$, the red to $\varepsilon=0.1$ and the black to $\varepsilon=0.15$. The dashed coloured lines show the value of $\varepsilon$ for the correspondingly coloured plot. Note that (a) the plots become flat for $M_R$ sufficiently large and (b) In this region, the value of these plots is approximately $\varepsilon$.}
\label{univ}
\end{figure}

Of important phenomenological consideration is the coupling of the standard--model fermions to the $Z$ boson. It has been shown by LEP, for the light fermions, that this coupling is universal with an accuracy at the per mille level \cite{PDG}. 
For the third generation quarks, where the picture is much less clear, we use the standard model as a guide. In the standard model there is no obvious reason for the coupling to be different for the third generation. In fact, any deviation from the universal value would be a signal of new physics. Also, the fact that flavour changing neutral currents are suppressed for the light generations of quarks suggests that some form of GIM mechanism must be at play. Detailed considerations are beyond the scope of this paper, but we note that this will be easier to achieve if universality of the gauge coupling applies to all three generations. For these reasons we will assume that the universality of the gauge coupling applies to all the standard--model fermions. 
This observation constrains the choices of parameters we may make in building a realistic model of standard--model fermion masses. To investigate this effect we introduce a bulk gauge sector\footnote{In the interest of simplicity when dealing with the fermion fields, we assume the gauge symmetry is $SU(2)_L\times U(1)_Y$.}, following the example of \cite{RL and MP}
\begin{eqnarray}\label{eqn:gaugeaction}
\mathcal{S}_{gauge}=-\frac{1}{4}\int d^4x\int^{r_2}_{r_1}dr\, \left(\frac{}{}a(r)-D b(r)\delta(r-r_2)\right)F_{\mu\nu}F^{\mu\nu}\nn\\+2b(r)F_{r\mu}F^{r\mu}-2b(r){\Omega^2}{}W^{a\mu} W^a_\mu\delta(r-r_1)\,,
\end{eqnarray}
where $F_{\mu\nu}$ represents the field--strength tensor of the gauge fields belonging to both the $SU(2)_L$ and $U(1)_Y$ groups and the functions $a(r)$ and $b(r)$ arise due to the curvature of the background, and are given by
\be
a(r)=1\,,\,\,b(r)=e^{2A}\,,
\ee
$Db(r_2)=r_2-\frac{1}{\varepsilon^2}$ is a UV kinetic term required for holographic renormalization of the gauge field 2-point functions, where $\varepsilon$ is a small parameter, and $\Omega$ is an IR VEV which controls electroweak symmetry breaking.
 Working in the unitary gauge ($W^a_r=0$) and the vector/axial--vector basis, we write
\begin{equation}
Z_\nu(q^2,r)=v(q^2,r)Z_\nu(q^2)\,,
\end{equation}
where $Z_\nu$ is the axial--vector gauge field.
We define $\partial_r v(q^2,r)\equiv\gamma(q^2,r)v(q^2,r)$, so that the equations of motion and IR boundary conditions for the axial--vector gauge field can be written
\be
\partial_r(b(r)\gamma(q^2,r))+b(r)(\gamma(q^2,r))^2+a(r)q^2=0\,,
\ee 
and
\be
\gamma(q^2,r_1)=\Omega^2\,.
\ee
The $Z$ boson is the zero mode of this field, defined by the expansion
\be
\gamma(q^2,r)=\gamma^0+q^2\gamma^1+\cdots\,,
\ee
for which the equation of motion reduces to
\be
\partial_r(b(r)\gamma^0)+b(r)(\gamma^0)^2=0\,,
\ee
while the IR boundary condition becomes
\be
\gamma^0=\Omega^2\,.
\ee
The coupling to fermions is introduced by modifying the covariant derivative in Eq.~(\ref{fermionaction}), such that
\be
\slashed{D}\rightarrow \slashed{\mathcal{D}}=\slashed{D}+\left(-i g\cos\theta_W T^3+i g^\prime\sin\theta_W Y \right) \slashed{Z}(q^2,r)\,,
\ee
where $\theta_W$ is the weak mixing angle, $T^3$ is the third generator of $SU(2)_L$, $Y$ is the hypercharge and $\slashed{D}=e^{\bar{M}}_A\Gamma^A D_{\bar{M}}$. The $SU(2)_L\times U(1)_Y$ symmetry of the standard--model then means that the left--handed term is identical for members of the same generation of quarks (or leptons) and the only difference in the coupling can arise due to a difference in the right--handed terms\footnote{One should note that up and down components are effected differently by the IR VEV. This fact means that differences in the coupling will eventually be generated for left handed up and down components. However, we expect this effect to be small and therefore ignore it here.}. For this reason we will concentrate on the right--handed fields only for this discussion. Factoring out the $r$ dependence and keeping only the zero modes as above, we find that the 4D coupling to the right--handed fermions is controlled by the integral
\be\label{int}
I=\int^{r_2}_{r_1}dr e^{3A(r)}v^0\left(f^0_{R}\right)^2\,,
\ee
where we work with normalized fields and
\be
v^0=e^{\int\gamma^0 dr}\,.
\ee
The only dependence on $\Omega$ comes from $v^0$. In the limit of large $\Omega$ this dependence drops out, as can be seen from the AdS solution
\be
v^0=1-\frac{\Omega^2}{2+\Omega^2}e^{-2r}\,,
\ee
Working in this limit we compute the integral Eq.~(\ref{int}). The results are presented in Figure \ref{univ} as a function of the bulk mass of the right handed fermion for $\delta\kappa=0$ and $\delta\kappa=2$. Figure \ref{univ} shows that the coupling becomes independent of the bulk mass above a certain value, which we call $M_0$. This is true at the per mille level. The value of $M_0$ is determined by $\delta\kappa$, as can be seen by comparing the two panels of Figure \ref{univ}, and this dependence is very approximately linear. Also, by considering the AdS result ($\delta\kappa=0$), we see that the dependence of the gauge coupling on the bulk mass under goes a transition around the classical value $M_{cl}$. This means that the region for which the coupling is independent is also the region in which the right handed field is localized in the UV. In fact  $M_0>M_{cl}$. This has important consequences for our approach to modeling the masses of the standard--model fermions. If we insist that the mass splitting between two fermions of the same generation is a result of the right handed fields having different bulk masses, then imposing the universality of the gauge coupling implies a lower bound on the bulk masses of these fields. Namely $M_R>M_0$ for each field. The results presented in Figure \ref{univ} also depend on the value of the small parameter $\varepsilon$. We present results for various values of $\varepsilon$ from which it can be seen that the value of $I$ in the region $M_R>M_0$ is approximately $\varepsilon$. This approximation is best for small $\varepsilon$ and increasing $\delta\kappa$.

\section{The $\hat{S}$ Parameter}

The results of the previous section have important consequences regarding the $\hat{S}$ parameter. This is because, if the coupling of the vector and axial-vector gauge fields to fermions is very different, the $\hat{S}$ parameter may receive large contributions. This can be checked by repeating the calculations of the previous section for the vector field and comparing the result. Specifically, we wish to compute
\be
I_v=\frac{\int^{r_2}_{r_1}dr e^{3A(r)}v_v^0\left(f^0_{R}\right)^2}{\left[\int^{r_2}_{r_1}dr e^{3A(r)}\left(f^0_R\right)^2\right]\left[\int^{r_2}_{r_1}dr v_v^0\left(1-D b(r_2)\delta(r-r_2)\right)v_v^0\right]^{-\frac{1}{2}}}\,,
\ee
where we choose to write the normalizations explicitly. The subscript $v$ denotes that we are now interested in the vector solutions and $v_v^0$ can be computed by following the previous procedure. Note that the equation of motion for $\gamma_v^0$ is the same as for $\gamma^0$ but the boundary conditions are different. Namely,
\be
\gamma_v^0(r_1)=0\,.
\ee
Solving the equation of motion subject to this boundary condition yields $\gamma^0_v=0$ such that $v_v^0=1$ for all $r$, hence $I_v$ reduces to
\be
I_v=\left[\int^{r_2}_{r_1}dr \left(1-D b(r_2)\delta(r-r_2)\right)\right]^{-\frac{1}{2}}=\varepsilon\,,
\ee
for $r_1=0$. Hence, if the corresponding result for the $Z$ is of order $\varepsilon$, the $\hat{S}$ parameter is likely to be small. By studying the results of the previous section we see that this is true only if $M_R>M_0$ and $\varepsilon$ is small. Note that similar considerations also apply to the left--handed fields. This is important as, requiring that the up and down components of the left --handed doublet couple in the same way does not ensure that this coupling is of the same order as the vector coupling. This requires $M_L<-M_0$. These results are true for the AdS case, as well as for when the additional sources of IR curvature are present. Interestingly, increasing the curvature in the IR actually improves this result. It is also interesting to note that this result is in agreement with both \cite{S1, S2} and \cite{DE and MP}.

\section{Top and Bottom}

We now turn our attention to using the formalism we have built up to model standard--model fermions. In order to make clear the effect of the IR curvature, we wish to build a model where the inter--generational splitting is controlled by $\lambda$, by each generation having a different Yukawa, and the intra--generational splitting is controlled by the bulk masses of the fields. We define $M_L$ to be the bulk mass of the left handed doublet containing the $t_L$ and $B_L$, $M_b$ the bulk mass of the $b_R$ and $M_t$ the bulk mass of the $t_R$. Also, we work in the limit of small $\lambda$. This can be justified since taking $\lambda$ large makes the mass splitting between the zero mode fermion and the first KK mode small. As no fermions heavier than the top have been observed we conclude that $\lambda$ must be small. In the context of this paper, we take small to mean $\lambda<1$ TeV. This seems reasonable since, in \cite{RL and MP}, the IR boundary (on which $\lambda$ is located) was introduced as a cut--off at the scale of confinement of the dual four dimensional field theory. This meant that the point $r_1$ coincided with an energy scale $\Lambda\sim\Lambda_{TC}\sim\mathcal{O}(1 {\rm TeV})$. Note that $\lambda$ is not the electroweak scale, which in \cite{RL and MP} would be $\Lambda_{TC}$. In fact one can view $\lambda$ as a dimensionless Yukawa $y$, multiplied by the scale $\Lambda_{TC}$.

\bs

With this kind of setup the hardest part of the mass hierarchy to explain is the masses of the top and bottom since this presents the largest intra--generational splitting, $m_t/m_b\sim 40$, and hence introduces the largest degree of tension upon the various parameters. For this reason we will focus only on top and bottom masses. Our approach is to look for values of $M_t$ and $M_b$ that give the correct values of the top and bottom masses\footnote{We take the $\bar{MS}$ masses for the top and bottom quoted in the PDG \cite{PDG}.} for various choices of the other parameters ($\lambda$, $M_L$, $r_2$ and $\delta\kappa$).

\begin{figure}[t]
\begin{center}
\includegraphics[]{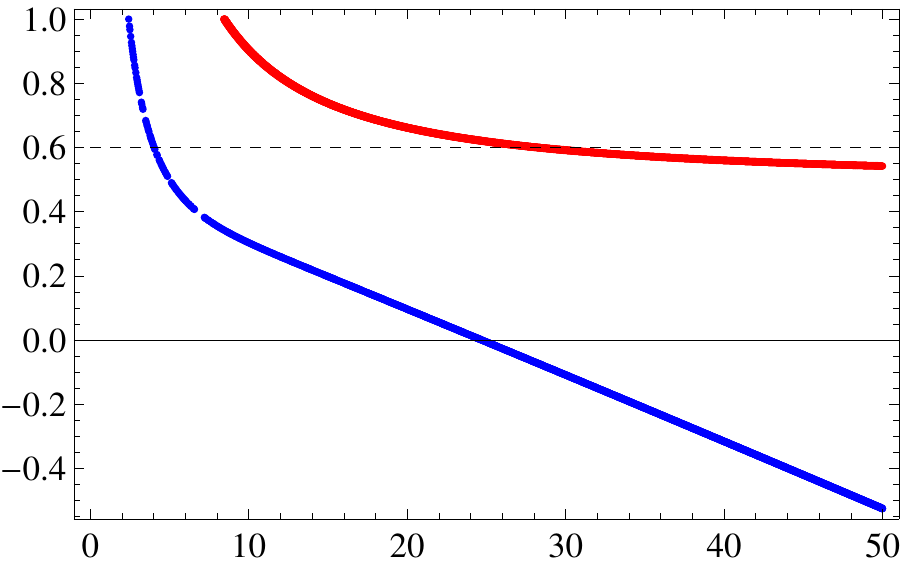}
\Text(-255,148)[]{\small{$M$}}
\Text(-20,0)[]{\small{$r_2$}}
\end{center}\caption{Plot of the the bulk masses of the top (blue curve) and the bottom (red curve) against the UV scale $r_2$ for $\delta\kappa=0$, given that the physical mass agrees with the $\bar{MS}$ value quoted in the PDG. The dashed line shows the lower bound on the bulk masses set by considering the universality of the gauge coupling. Note that only small values of the UV scale give a fit which is consistent with this bound.}
\label{littletopmodel}
\end{figure}

\bs

The first scenario we will consider is for $\lambda=0.8$ TeV, $M_L=-1.001/2$ and $\delta\kappa=0$. This corresponds to AdS, so we need not consider the parameters $\Delta$ or $r_*$ when making this assignment for $\delta\kappa$. We also set $\kappa_1\equiv\kappa=1$. Figure \ref{littletopmodel} then shows how $M_t$ and $M_b$ vary as the UV scale $r_2$ is changed. The dashed line in the plot shows the lower bound obtained by considering the universality of the gauge coupling, as discussed in the previous section. From this we see that phenomenologically viable solutions only exist for very small values of the UV scale, $r_2<5$. Notice that our choice of $M_L$ means that the left handed doublet is localized in the UV, while the bound set by the universality of the gauge coupling ensures that the $t_R$ and $b_R$ are also.

\bs

\begin{figure}[t]
\begin{center}
\includegraphics[]{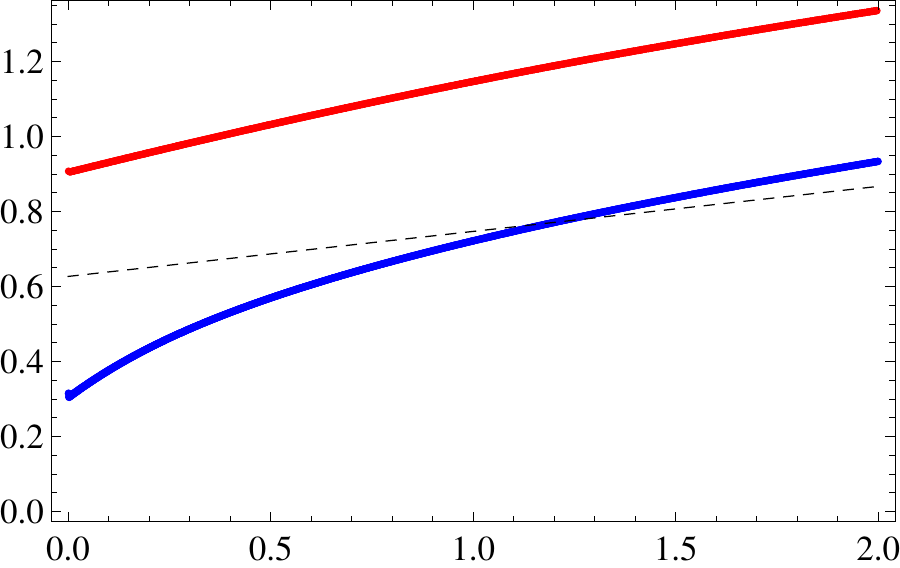}
\Text(-255,155)[]{\small{$M$}}
\Text(-20,0)[]{\small{$\delta\kappa$}}
\end{center}\caption{Plot of the bulk masses of the top (blue) and bottom (red) against $\delta\kappa$, given the physical masses of the top and bottom, for $\lambda/{\rm TeV}=0.8$, $r_*=2.5$, $r_2=10$ and $M_L=-1.001/2$. The black dashed line shows the lower bound on the bulk mass for both $t_R$ and $b_R$, obtained by considering the universality of the gauge coupling. Note that a consistent fit is only possible for $\delta\kappa>1.2$.}
\label{plotdk}
\end{figure}

 We are finally ready to exemplify the main element of novelty in our approach.
We want to gauge what effect turning on additional sources of curvature in the IR, by taking $\delta\kappa\neq 0$, can yield. To do so we begin by fixing $\lambda = 0.8$ TeV, $\kappa_1=1$, $r_*=2.5$, $r_2=10$, $M_L=-1.001/2$ and assume that $\Delta$ is large so that the curvature for $r<r_*$ is controlled by $\delta\kappa$ in Eq.~(\ref{dk}). Our results are presented in Figure \ref{plotdk}, where again the dashed line shows the lower bound set by the universality of the gauge coupling. We see in this case that viable solutions only exist for large IR curvatures, $\delta\kappa>1.2$.
 
\bs

\begin{figure}[t]
\begin{center}
\includegraphics[]{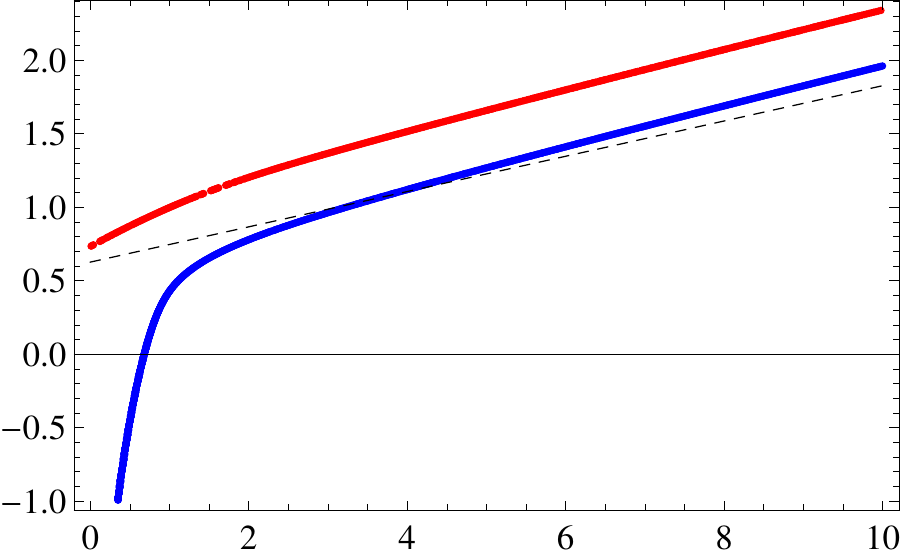}
\Text(-255,155)[]{\small{$M$}}
\Text(-20,0)[]{\small{$\delta\kappa$}}
\end{center}\caption{Plot of the bulk masses of the top (blue) and bottom (red) against $\delta\kappa$, given the physical masses of the top and bottom, for $\lambda/{\rm TeV}=0.246$, $r_*=2.5$, $r_2=10$ and $M_L=-0.55$. In this case large values of $\delta\kappa$ must be taken to find a consistent fit.}
\label{lambdasmall}
\end{figure}

Finally, we consider a smaller Yukawa $\lambda=0.246$ TeV, setting $M_L=-0.55$ and $r_2=10$. Repeating our analysis for this case then yields Figure \ref{lambdasmall}. In this case the IR curvature needs taking much larger before the bound on the bulk masses can be satisfied.

\bs

Considering these three examples, we extract some conclusions about the role of the IR curvature in producing the desired four dimensional mass spectrum. In the case when $\delta\kappa=0$ (AdS), producing the correct inter--generational mass splitting, while also satisfying the constraint set by the universality of the gauge coupling, requires the UV scale $r_2$ to be taken unnaturally small. This problem is overcome when additional sources of IR curvature are turned on. It also turns out that the values of $M_t$ and $M_b$ that can be taken are somewhat close to each other, which is a desirable feature. The result is that increasing the curvature in the IR has a positive effect on our attempts to reproduce the intra--generational mass splitting. We also note that, due to the way we choose set up our model, if we were to consider modeling the masses of the charm and strange (for which a bound set by the universality of the gauge coupling is well known) we would find very similar results to those we found for the top and bottom. This is because $m_c/m_s\sim 40$ also. Also, to find the results for the charm, we would take the results for the top and re-scale $\lambda$. Therefore, the resulting plot would be somewhat identical to Figures \ref{littletopmodel} and \ref{plotdk}.

\bs

In order to understand this behaviour we note that the mass of a given fermion is controlled by how it is localized in the bulk. In AdS ($\delta\kappa=0$), the fermions are classical for $M_L=-1/2$, $M_R=1/2$ and the wavefunction is flat in the bulk. If the magnitude of the bulk masses is increased the fermions localize in the UV. This suppresses the mass since the overlap with the IR boundary is small. Decreasing the magnitude of the bulk masses localizes the fermions in the IR, generating a large mass. In the deformed background the situation is more complicated. When the bulk masses are $M_L=-1/2$, $M_R=1/2$ the fermions are classical {\it in the UV}, by which we mean the wavefunction is flat for $r>r_*$, and localized {\it in the IR}. The fermion can also be classical in the IR and localized in the UV. This occurs for $M_L=-\kappa_0/2$, $M_R=\kappa_0/2$. Intermediate solutions, where the wavefunction is peaked both in the IR and UV are also possible. It is this structure that is ultimately responsible for the behaviour we see. Fixing $M_L$, such that the left--handed doublet is mildly localized in the UV (i.e $M_L$ lies close to the UV classical solution), and increasing the curvature in the IR makes the wave function of the left--handed doublet more peaked, and eventually localized, in the IR.  It is the overlap of the wavefunction of the left-- and right--handed fermions and the IR boundary that determines the physical mass. Therefore, in order to obtain a fixed 4D mass whilst increasing $\delta\kappa$, $M_R$ must also be increased to compensate. Finally, $M_t$ is more sensitive to changes in IR curvature than $M_0$ (the bound). Hence, the bound can eventually be satisfied by increasing $\delta\kappa$.

\section{Discussion}

For illustrational purposes, in our analysis of the mass hierarchy of the top and bottom we considered the case where they share the same Yukawa. This allowed us to see the dependence of the masses upon the other parameters in the model. In particular we focused on the interplay between the curvature in the IR and the bulk masses of the right--handed fields. Using the universality of the gauge coupling to fermions as a phenomenological constraint we saw that for $\lambda=v_W$ viable solutions only exist for large values of IR curvature $\delta\kappa>3.5$. Taking $\lambda$ larger than $v_W$ yields viable solutions for smaller values of the IR curvature and this behaviour is controlled by how the fermions localize in the bulk.

\bs

Having paid much attention to the top and bottom, we feel it is necessary to also comment on the other standard--model fermions. Since the UV scale and curvatures are properties of the background geometry, the values we choose for the top and bottom must be kept universal. This leaves only the Yukawas and bulk masses that can be varied in order to yield the correct spectrum for the other fermions. The intra--generational hierarchy between the charm and strange is similar to that of the top and bottom, so it seems reasonable to change the Yukawa in such a way that $m_t\rightarrow m_c$. A slight adjustment of the bulk mass of the $s_R$ should then be all that is required to yield the correct spectrum. As such, we do not expect the bound coming from the universality of the gauge coupling to present much of a problem here. Much trickier is the fit for the up and down, since $m_u/m_d<1$.

\bs

In fitting the fermion masses for the other generations, one should avoid changing $M_L$ as this has the potential to cause problems for the universality of the gauge coupling. We have ignored the bounds on $M_L$ because they do not matter for a single generation, but if $M_L$ is to vary between generations then similar limits exist as those for the bulk masses of the right--handed fields (in AdS the bound is $M_L<-0.6$ for each generation).

\bs

It is interesting to note that this setup is starkly different to the standard--model, where each flavour of fermion has its own Yukawa. It is possible to modify our approach to make this model look more standard--model like, by simply introducing more Yukawas and this would make the job of fitting the fermion masses somewhat trivial. In this case, changing the curvature in the IR simply introduces another parameter that can be chosen so as to find a satisfactory fit. In fact, the generational structure we consider is more akin to extended technicolor.
We do not suggest that what we do here is enough to model extended technicolor correctly. More careful analysis is required to see if this is indeed possible, which is beyond the scope of this paper. It is interesting to note though that this approach gives a way to explain the intra--generational hierarchies of the standard--model fermions purely in terms of bulk masses and the curvature of the background, i.e. renormalization group effects. In this context the effect of these parameters could be seen as corrections to the fermion masses arising from strong dynamics to which perturbative calculations are insensitive. As such, this could be a very useful asset for model building.

\section*{Acknowledgments}

We wish to thank Maurizio Piai for useful discussions. The work of RL is supported by STFC Doctoral Training Grant ST/I506037/1.

\end{document}